\documentclass[12pt]{article}
\usepackage{amsfonts,amsmath,amssymb,epsf}

\newcommand{\tr}{{\rm tr\,}}

\newcommand{\bR}{\mathbf{R}}

\newcommand{\nn}{\nonumber}
\newcommand{\commt}[2]{\left[#1, #2 \right] }

\newcommand{\ph}[1]{\phantom{#1}}

\newcommand{\ul}[1]{\underline{#1}}
\newcommand{\thcommt}[3]{\left[#1, #2, #3 \right] }

\newcommand{\fijk}[3]{f^{#1 #2}_{\phantom{#1 #2} #3}}
\newcommand{\fABCD}[4]{f^{#1 #2 #3}_{\phantom{#1 #2 #3} #4}}
\newcommand{\FAB}[4]{\tilde{F}^{\phantom{#1 #2} #3}_{#1 #2 \phantom{#3} #4}}
\newcommand{\AAB}[3]{\tilde{A}^{\phantom{#1} #2}_{#1 \phantom{#2} #3}}

\numberwithin{equation}{section}
\allowdisplaybreaks

\renewcommand{\title}[1]{\vbox{\center\LARGE{#1}}\vspace{5mm}}
\renewcommand{\author}[1]{\vbox{\center#1}\vspace{5mm}}
\newcommand{\address}[1]{\vbox{\center\em#1}}
\newcommand{\email}[1]{\vbox{\center\tt#1}\vspace{5mm}}

\begin{document}

\begin{titlepage}
\rightline{\today}

\vspace{30mm}

\title{
 Branes from a non-Abelian (2,0) tensor multiplet with 3-algebra
}

\author{Shoichi Kawamoto$^1$, Tomohisa Takimi$^2$ and Dan Tomino$^3$}

\address{$^1$Department of Physics, National Taiwan Normal University,
  Taiwan\\
\medskip
$^2$
Department of Physics, National Taiwan University,
  Taiwan\\
\medskip
$^3$
National Center for Theoretical Science (NCTS), Hsinchu, Taiwan
}

\email{kawamoto@ntnu.edu.tw, tomo.takimi@gmail.com, tomino@phys.cts.nthu.edu.tw}

\abstract{
In this paper, we study the equations of motion for non-Abelian 
$\mathcal{N}=(2,0)$
tensor multiplets in six dimensions,
which were recently proposed by Lambert and
Papageorgakis. 
Some equations are regarded as
constraint equations. 
We employ a loop extension of the Lorentzian
three-algebra (3-algebra) and examine the equations of motion around
various solutions of the constraint equations. 
The resultant equations
take forms that allow Lagrangian descriptions. 
We find various ($5+d$)-dimensional Lagrangians and
investigate the relation between them from the viewpoint of M-theory
duality.
}

\end{titlepage}

\section{Introduction}
\label{sec:intro}

Since its discovery, M-theory has been intensively studied from
various viewpoints, such as string duality and its applications to
supersymmetric gauge theories. 
Despite extensive study since the
1990s, the basic properties of M-theory, including its fundamental
degrees of freedom, still remain mysterious. 
However, there are a
number of aspects of M-theory which have been clarified. 
For instance,
the low-energy limit of this theory is  11-dimensional
supergravity and, at least in the long-wavelength approximation, it
accommodates two kinds of extended object, M-theory two-branes (M2-branes)
and five-branes (M5-branes), which couple to the three-form gauge
fields in 11-dimensional supergravity.

Similarly, there are still numerous aspects of M-branes to unveil.
In particular, the world-volume description of multiple M-branes is
interesting in the context of the AdS/CFT correspondence and, more
importantly, M-branes are believed to be described by novel interacting
superconformal field theories, such as a three-dimensional 
$\mathcal{N}=8$
superconformal field theory for M2-branes and a six-dimensional
superconformal field theory of $(2,0)$ tensor multiplets for
M5-branes. 
However, the formulation of the theory at the fundamental
level poses a difficult problem which has remained unresolved for a
long time.

Recently, a world-volume description of multiple M2-branes using a new
kind of symmetry structure, the so-called Lie 3-algebra, was proposed
independently by Bagger and
Lambert~\cite{Bagger:2006sk,Bagger:2007jr},
and Gustavsson~\cite{Gustavsson:2007vu} (BLG).
Since then,
significant progress has been made in understanding the BLG theory and
3-algebra itself.
In the course of the study of 3-algebra, it was
first conjectured~\cite{Ho:2008bn} and later proved\footnote{%
We have been informed of the paper~\cite{Nagy} which
claims to provide the first proof to this
conjecture.}~\cite{Papadopoulos:2008sk,Gauntlett:2008uf}
that the only
finite-dimensional Lie 3-algebras with a positive definite metric are
the trivial one, $\mathcal{A}_4$, and the direct sum of these algebras.
Multiple M2-branes can also be reformulated in the context of
the double Chern-Simons theory with the usual Lie group
symmetry~\cite{VanRaamsdonk:2008ft,Aharony:2008ug}\footnote{%
There is recent research on the hidden maximal supersymmetry of the
Chern-Simons-matter theory with less than $\mathcal{N}=8$
supersymmetry; for example, see \cite{Gustavsson:2009pm}.},
and thus the 3-algebra might not be indispensable in describing
multiple M2-branes. 
On the contrary, an $\mathcal{N}=6$ Chern-Simons-matter
theory in three dimensions, the so-called ABJM theory, also revealed
hidden 3-algebraic structures~\cite{Bagger:2008se},
and S-matrix analysis of
three-dimensional (3D) $\mathcal{N}=8$ on-shell supermultiplets suggested
that interaction has 3-algebraic structures~\cite{Huang:2010rn}.
Therefore, 3-algebra
could play a crucial role in the analysis of the dynamics of M-theory,
but a better understanding of its attributes is necessary.
Moreover,
since we can formulate a system of an infinite number of M2-branes
which can be condensed to a single M5-brane by means of
infinite-dimensional Lie 3-algebra\footnote{%
The Lie 3-algebra resulting in a Nambu-bracket structure was studied
in detail by \cite{Ho:2008nn,Ho:2008ve,Ho:2009mi,Chu:2008qv}.
The approach of employing an infinite-dimensional
algebra is independent from the no-go theorem mentioned
above~\cite{Ho:2008bn,Papadopoulos:2008sk,Gauntlett:2008uf}.},
it would also be possible to use
3-algebra to formulate a system of interacting M5-branes manifesting
$(2,0)$ supersymmetry.

Recently, Lambert and Papageorgakis~\cite{Lambert:2010wm}
 proposed a set of on-shell
supersymmetry transformations for non-Abelian $(2, 0)$ tensor multiplets
in six dimensions, using a 3-algebraic structure. 
This approach immediately invokes the following question: to what
extent is it related to multiple M5-branes systems?

The supersymmetry transformations proposed by Lambert and
Papageorgakis were as follows:
\begin{align}
\label{SUSY_X}
\delta X^I_A=& i\bar{\epsilon}\Gamma^I\Psi_A \,, 
\nonumber \\
\delta \Psi_A=& \Gamma^{\mu}\Gamma^ID_{\mu}X_A^I\epsilon
+\frac{1}{3!}\frac{1}{2} \Gamma_{\mu\nu\lambda}H^{\mu\nu\lambda}_A \epsilon
-\frac{1}{2}\Gamma_{\lambda}\Gamma^{IJ}C^{\lambda}_BX^I_CX^J_D{f^{CDB}}_A\epsilon
\,,
\nonumber \\
\delta H_{\mu\nu\lambda\;A}=& 3i\bar{\epsilon}\Gamma_{[\mu\nu}D_{\lambda]}\Psi_A
+i\bar{\epsilon}\Gamma^I\Gamma_{\mu\nu\lambda\kappa}C_B^{\kappa}X^I_C\Psi_D{f^{CDB}}_A 
\,,
\nonumber \\
\delta \tilde{A}_{\mu\; A}^B=&
i\bar{\epsilon}\Gamma_{\mu\lambda}C_C^{\lambda}\Psi_D{f^{CDB}}_A \,, 
\nonumber \\
\delta C^{\mu}_A=&0. 
\end{align}
The three-form $H_{\mu\nu\lambda A}$ satisfies the linear self-duality
condition:
\begin{align}
H_{\mu\nu\lambda
 A}=\frac{1}{3!}\epsilon_{\mu\nu\lambda\tau\sigma\rho}{H^{\tau\sigma\rho}}_A. 
\label{self-dual}
\end{align}
The scalar fields $X^I$, the fermions $\Psi$, and the self-dual field
$H_{\mu\nu\rho}$ form a $(2, 0)$ tensor multiplet in six dimensions.
The two world-volume vectors $\tilde{A}_{\mu\; A}^B$ and $C^{\mu}_A$
are new and play an important role in the introduction of a
non-Abelian structure for the tensor multiplets.
The gauge covariant
derivative $D_{\mu}$ is defined by $D_{\mu}X^I_A=\partial_{\mu}X^I_A
- \tilde{A}_{\mu\;A}^BX^I_B$.
(See Appendix \ref{sec:DandF} and 
\ref{sec:assum_gauge} for more details.)
The Greek indices such as $\mu$, $\nu$, represent
the world-volume directions and run from $0$ to $5$.
$I,J$ indices are for the transverse
directions to the world-volume, $I,J=6, \cdots 10$.
$A,B,\cdots$ denote the gauge indices of the
3-algebra symmetry.

In order for the above transformations to result in an on-shell
supersymmetry on the fields, the equations of motion and the
constraints of the fields are necessary for their closure. These
equations were derived in  \cite{Lambert:2010wm} and are\footnote{%
We find that the sign for the last term in (\ref{feq1})
should be ``$+$''
and also the sign in (\ref{feq4}) is corrected to be ``$+$.''}:
\begin{align}
0=& (D^2X^I)_A-\frac{i}{2}\bar{\Psi}_CC_B^{\nu}\Gamma_{\nu}\Gamma^I\Psi_D{f^{CDB}}_A
+C^{\nu}_BC_{\nu G}X^J_CX^J_EX^I_F{f^{EFG}}_D{f^{CDB}}_A \,, 
\label{feq1} \\
0=&  \left( D_{[\mu}H_{\nu\lambda\rho]}\right)_A
+\frac{1}{4}\epsilon_{\mu\nu\lambda\rho\sigma\tau} C^{\sigma}_BX^I_CD^{\tau}X^I_D{f^{CDB}}_A
+\frac{i}{8}\epsilon_{\mu\nu\lambda\rho\sigma\tau}
C^{\sigma}_B\bar{\Psi}_C\Gamma^{\tau}\Psi_D{f^{CDB}}_A \,,   
\label{feq2} \\
0=&   \Gamma^{\mu} (D_{\mu}\Psi)_A
+X^I_CC^{\nu}_B\Gamma_{\nu}\Gamma^I\Psi_D{f^{CDB}}_A \,,  
\label{feq3} \\
0=&  \tilde{F}_{\mu\nu\;\; A}^{\;\;B}  +C^{\lambda}_CH_{\mu\nu\lambda
  D}{f^{CDB}}_A \,, 
\label{feq4} \\
0=&  D_{\mu}C^{\nu}_A =C^{\mu}_CC^{\nu}_D{f^{CDB}}_A \,, 
\label{feq5} \\
0=&  C^{\rho}_C (D_{\rho}X^I)_D {f^{CDB}}_A
=C^{\rho}_C (D_{\rho}\Psi)_D{f^{CDB}}_A =
C^{\rho}_C (D_{\rho}H_{\mu\nu\lambda})_ D{f^{CDB}}_A \,,
\label{feq6}
\end{align}
where $\tilde{F}_{\mu\nu\;\;A}^{\;\;B}={[D_{\mu},D_{\nu}]^B}_A$.

At first, these equations seem suitable for describing multiple
M5-branes because of their non-Abelian structure introduced by
3-algebra. 
However, it was pointed out in \cite{Lambert:2010wm} that
when a simple Lorentzian 3-algebra or the $\mathcal{A}_4$-algebra is
adopted, the equations describe multiple 4-branes and not
5-branes. 
With the use of these algebras, one of the world-volume
directions is eliminated through equation (\ref{feq6}), and the resulting
equations of motion
can describe the five-dimensional (5D) supersymmetric Yang-Mills (SYM)
theory. 
In other words, we ultimately describe D4-branes rather than
M5-branes.
This is in many aspects similar to the "M2 to D2" scenario
\cite{Mukhi:2008ux, Ho:2008ei,  Ezhuthachan:2008ch}
proposed for the BLG theory, especially for the Lorentzian
case.
One may therefore wonder whether the non-Abelian tensor
multiplet can describe more than the dynamics of D4-branes and
5-branes of M-theory.
In this paper, we address this question
by examining equations 
(\ref{feq1})--(\ref{feq6}) based on a 3-algebra of infinite
dimensions. \\

Through a systematic study of the fundamental identity, a class of
Lorentzian 3-algebras has been proposed in \cite{Ho:2009nk},
which includes the
simplest Lorentzian 3-algebra used in \cite{Lambert:2010wm, Ho:2008ei}
as a special case.
In this
paper, we use only one of the proposed 3-algebras:
\begin{align}
[u^0,u^a,u^b]=&  0 \,, 
\label{algebra1}\\
{} [ u^0, u^a,T^{(i\vec{m})} ]=& m^aT^{(i\vec{m})}  \,,
\label{algebra2}\\
{}[u^0,T^{(i\vec{m})},T^{(j\vec{n})}]=& m^a\delta^{ij}\delta^{\vec{m}+\vec{n}}
u^{\ul{a}}+i{f^{ij}}_kT^{(k,\vec{m}+\vec{n})}  \,,
\label{algebra3}\\
{}[u^a,T^{(i\vec{m})},T^{(j\vec{n})}]=& -m^a\delta^{ij}\delta^{\vec{m}+\vec{n}}
u^{\ul{0}}  \,,
\label{algebra4}\\
{}[T^{(i\vec{m})},T^{(j\vec{n})},T^{(k\vec{l})}]=&
-if^{ijk}\delta^{\vec{m}+\vec{n}+\vec{l}}\; u^{\ul{0}} \,,
\label{algebra5}
\end{align}
with inner products of
\begin{align}
\label{algebra_inner_prod}
\langle T^{(i\vec{m})}, T^{(j\vec{n})} \rangle = \delta^{ij}
\delta^{\vec{m}+\vec{n}}, \quad
\langle u^a, u_b \rangle = \delta^{a}_{b},
\quad \langle u^0, u_0 \rangle = 1 .
\end{align}
Other combinations of inner products are all equal to zero.
It should
be noted that the three-bracket is defined as 
$[ T^A , T^B , T^C ] =i f^{ABC}{}_D T^D$,
and $T^A$ as $T^A=\{u^0, u^a,
u^{\underline{0}},  u^{\underline{a}}, T^{(i\vec{m})}\}$.
Namely the index such as $A$ represents different kinds of generator;
the Euclidean part $i,j,\cdots$ accompanied with a $d$-dimensional
numerical vector such as $\vec{m}$, and the center part $0,a,\ul{0}$
and $\ul{a}$.
$d$ denotes an integer and
the indices $a$ and $b$ run from $1$ to $d$.
$u^{\underline{0}}$ and
$u^{\underline{a}}$ are center elements of the algebra.
Furthermore, ${f^{ij}}_k$ is the structure constant of a Lie algebra,
$[T^{(i\vec{m})}, T^{(j\vec{n})}] =
if^{ij}{}_{k}T^{(k,\vec{m}+\vec{n})}$. 
Lastly, $\vec{m}$ is a
$d$-dimensional vector whose components are $m^a$ ($a=1,\cdots,d$) and
are considered to be integers. 
A more detailed description of this
3-algebra is provided in Appendix \ref{sec:algebra}.

Applying this infinite-dimensional 3-algebra to the BLG theory, we
can obtain the SYM theory on a torus \cite{Ho:2009nk}.
Specifically, vector $\vec{m}$
turns out to play the role of a Kaluza-Klein (KK) momentum vector on
this torus. 
Thanks to the KK-momentum, new world-volume directions can
be introduced and the BLG theory based on the 3-algebra is able to
describe more than two-dimensional branes.
The U-duality relation of
D-branes wrapping on the torus has been studied in \cite{Kobo:2009gz}.
We therefore
expect to get not only 4-branes but also higher-dimensional-branes
using the 3-algebraic field equations
(\ref{feq1})--(\ref{feq6}).\\

In the following sections, we analyze these field equations with the
use of 3-algebra
(\ref{algebra1})--(\ref{algebra_inner_prod}).
We will see that it is important to
understand the constraints of $C^{\mu}$, \textit{i.e.}, (\ref{feq5}),
in order to obtain the effective action of branes from the field
equations.
In section \ref{sec:gen_case}, we consider the most generic case of
$C^{\mu}=C_{0}^{\mu}u^{0}+C^{\mu}_au^a+C^{\mu}_{(i\vec{m})}T^{(i\vec{m})}$.
In this case, we can describe the effective action of $(4+d)$-brane
wrapping on a torus $T^{d}$. More specifically, the brane is described
in target space ${\bf R}^{1,5}\times T^d \times {\bf R}^{4-d}$.
In section \ref{sec:C0_zero}, we consider the case of
$C^{\mu}=C^{\mu}_au^a$.
For $C^{\mu}_0=0$, we can describe the
effective action of 5-branes wrapping on the torus $T^d$, including
non-covariant massive vector bosons.
A SYM type action is recovered when $d=1$.
Regardless of the non-Abelian structure introduced by
3-algebra, the resulting 5-branes theory turns out to be
Abelian.
These 5-branes are in the target space ${\bf R}^{1,5-d} \times
T^d\times {\bf R}^{4}$.
In section \ref{sec:vanishing_C},
we consider the case of
$C^{\mu}_A=0$ and recover the second-order Pasti-Sorokin-Tonin
(PST~\cite{Pasti:1997gx}) type action of NS5-branes~\cite{Bandos:2000az}.
The target space of the 5-branes is
reduced to ${\bf R}^{1,9-d}\times M_{d+1}$, where $M_{d+1}$ is a
$(d+1)$-dimensional manifold.
Namely, the M-theory is compactified on $M_{d+1}$.
In section \ref{sec:5-branes},
we concentrate on the case of $d=1$ under different
$C^{\mu}$, and several 5-branes obtained in the preceding sections
and their relations are investigated.
The latter can be identified as
5-branes in a type I{}IA/I{}IB string theory.
We find that the S-dual
relation between D5-branes and NS5-branes is naturally
realized.
Finally, section \ref{sec:conclusion} is devoted to discussion and concludes
with a summary.
We also comment on recent research on the 5D maximally
supersymmetric Yang-Mills (MSYM) theory proposed by
\cite{Lambert:2010iw, Douglas:2010iu},
from our viewpoint.
Three appendices provide a summary of notations and
conventions, and supplementary discussions.


\section{Analysis of the equations of motion}
\label{sec:eoms_analysis}

In this section, we examine the equations of motion
(\ref{feq1})--(\ref{feq6}), with the infinite dimensional extension of
Lorentzian type 3-algebra (\ref{algebra1})--(\ref{algebra_inner_prod}).
The basic properties of this 3-algebra are summarized in Appendix
\ref{sec:algebra}.
The equation (\ref{feq5}) is considered
as the constraints for $C_A^\mu$.
We thus start with solving these constraints,
and then move on to the examination of the rest.

\subsection{The gauge fields and the constraints for $C_A^\mu$ fields}
\label{sec:constraints}

First, in this paper
we assume that the gauge field $\AAB{\mu}{B}{A}$ is accompanied
with the structure constant, 
$\AAB{\mu}{B}{A} \equiv A_{\mu\; CD} f^{CDB}{}_A$.
It should be noted that though this is a requirement for the BLG model
(M2-brane case)
due to Chern-Simons term, it is not necessary in this case.
But we also adopt this definition here, since it guarantees that the
covariant derivative acts on the three-bracket as a derivation.
Because of the limited form of the structure constant, some components
of the gauge field vanish;
\begin{align}
\AAB{\mu}{0}{\ul{0}}=&
\AAB{\mu}{a}{\ul{b}}=
\AAB{\mu}{A}{0}=
\AAB{\mu}{A}{a}=
\AAB{\mu}{\ul{0}}{A}=
\AAB{\mu}{\ul{a}}{A}=0 \,.
\end{align}
This fact simplifies our analysis.
The nonzero components of the gauge fields are summarized in Appendix
\ref{sec:assum_gauge}.

Next, we consider the first equation of (\ref{feq5}):
\begin{align}
  \label{eq:DC_const1}
  \left( D_\mu C^\nu \right)_A =& 0 \,.
\end{align}
For $u^0$ and $u^a$ components, this condition immediately means
\begin{align}
  \partial_\mu C_0^\nu = \partial_\mu C_a^\nu =0 
\end{align}
for arbitrary $\mu$ and $\nu$.
Therefore, 
$C_0^\mu$ and $C_a^\mu$ are constants.
As for $u^{\ul{0}}$ and $u^{\ul{a}}$ components,
since in all the other places $C^\mu_A$ always appear with the structure
constant $f^{BCD}{}_A$,
$C^\mu_{\ul{0}}$ and $C^\mu_{\ul{a}}$ show up only in these constraint
equations:
\begin{align}
  \partial_\mu C_{\ul{0}}^\nu 
=& \AAB{\mu}{a}{\ul{0}} C_a^\nu
+\AAB{\mu}{(i\vec{m})}{\ul{0}} C_{(i\vec{m})}^\nu
 \,,
\nn\\
  \partial_\mu C_{\ul{a}}^\nu 
=& \AAB{\mu}{0}{\ul{a}} C_0^\nu
+\AAB{\mu}{(i\vec{m})}{\ul{a}} C_{(i\vec{m})}^\nu
 \,.
\label{eq:DC_ghost1}
\end{align}
Therefore, $C_{\ul{0}}^\nu $ and $C_{\ul{a}}^\nu $ are completely
determined by these two equations.
As we will argue in section \ref{sec:ghost_decoupling},
 the components associated with the center elements of
the 3-algebra are regarded as the ghost fields.
Therefore, these conditions 
imply that the ghost fields are
excited by the physical fields.
To avoid it, we will impose the condition that the ghost fields
stay constant:
\begin{align}
  \partial_\mu C_{\ul{0}}^\nu 
=  \partial_\mu C_{\ul{a}}^\nu 
=0  
\end{align}
in 
time evolution.

There are also constraints imposed on
$C^{\nu}_{(i\vec{m})}$, which are 
\begin{equation}
0 =  (\tilde{D}_{\mu}C^{\nu})_{(i\vec{m})}
- \AAB{\mu}{0}{(i\vec{m})} C^{\nu}_{0}
- \AAB{\mu}{a}{(i\vec{m})} C^{\nu}_{a} \label{Cim constraint}
\end{equation}
where 
the covariant derivative $\tilde{D}_{\mu}$ 
is given by (\ref{eq:def_Tilde_D}):
\begin{equation}
(\tilde{D}_{\mu}C^{\nu})_{(i\vec{m})}
= 
\partial_\mu C^{\nu}_{(i\vec{m})}
-\tilde{A}^{(i\vec{m})}_{\mu\, (i\vec{m})} C^{\nu}_{(i\vec{m})} 
+i \commt{A_\mu}{C^{\nu}}_{(i\vec{m})} \,. 
\end{equation}

Finally, we examine the second equation of (\ref{feq5}):
\begin{align}
\label{eq:CC_const1}
    \thcommt{C^\mu}{C^\nu}{T^B}_{A} =& 0 \,.
\end{align}
For the various combinations of
$A,B=\{(i\vec{m}), 0, a, \underline{0}, \underline{a} \}$,
we have
\begin{align}
    \label{eq:CC_const2}
0=&
f^{ki}{}_j  C_0^{[\mu} C_{(k,\vec{m}-\vec{n})}^{\nu]} 
= m^b C_0^{[\mu} C_b^{\nu]}
\nn\\=&
   m^b C_b^{[\mu} C_{(i\vec{m})}^{\nu]} - \commt{C^\mu}{C^\nu}_{(i\vec{m})}
=   m^a C_0^{[\mu} C_{(i\vec{m})}^{\nu]}
=C_{(k\vec{\ell})}^{[\mu} C_{(k,-\vec{\ell})}^{\nu]} \,,
\end{align}
where in the first equation $\vec{m} \neq \vec{n}$ , and
the repeated indices in 
a single term
(here denoted as $b$, $k$ and the vector $\vec{\ell}$)
are all implicitly summed over, and we will 
use this contraction rule throughout this paper.
$f^{ij}{}_k$ is the structure
constant for a conventional Lie algebra which is embedded into our
3-algebra,
and
\begin{align}
\label{eq:def_2-commt}
    \commt{\phi}{\varphi}_{(i\vec{m})}
\equiv & i f^{jk}{}_i \phi_{(j\vec{n})} \varphi_{(k,\vec{m}-\vec{n})} \,.
\end{align}
In (\ref{eq:CC_const2}), the last condition is trivial.
If only one direction of $C^\mu$ is non-vanishing,
all the constraints are trivially satisfied.
Otherwise, these constraints give restriction on $C$ fields.
To solve the constraints, we consider the following possibilities:
\begin{enumerate}
\item All $C^\mu_0$, $C^\mu_a$ and $C^\mu_{(i\vec{m})}$ are nonzero:

In this most generic case,
to satisfy the constraints there have to be the relations;
$C^\mu_a \propto C^\mu_0$ and $C^\mu_{(i\vec{m})} \propto
  C^\mu_0$.
Therefore, we have the following conditions,

\begin{align}
  \label{eq:gen_C_cond1}
  C^\mu_a = v_a C^\mu_0 \,,
\qquad
C^\mu_{(i\vec{m})} = v_{(i\vec{m})} C^\mu_0 \,,
\end{align}
where $v_a$ and $v_{(i\vec{m})}$ are chosen in common for all $\mu$.
Note that $v_{(i\vec{m})}$ are commuting each other with respect to
the commutator (\ref{eq:def_2-commt}),
 $\commt{v}{v}_{(i\vec{m})}=0$.
\vspace{1em}

\item $C^\mu_0 \neq 0$ but $C^\mu_a = C^\mu_{(i\vec{m})}=0$:

This case is included in the previous case with $v_a=v_{(i\vec{m})}=0$.
\vspace{1em}

\item $C^\mu_0=C^\mu_{(i\vec{m})}=0$ and $C^\mu_a \neq 0$:

In this case, $C^\mu_a$ would take arbitrary constant values
which are not necessarily proportional to one another.

\vspace{1em}

\item All $C^\mu_0=C^\mu_a=C^\mu_{(i\vec{m})}=0$ case:

In this case, the non-Abelian interactions are almost turned off.
\end{enumerate}

We do not try to exhaust all the possibilities but look at interesting
cases.
In the following subsections, we are going to investigate the
equations of motion for each of the above cases.
(The case 2 is included in the case 1.)
Before going to the analysis, we discuss the decoupling of the
ghost fields.

\subsection{Decoupling of the ghost fields}
\label{sec:ghost_decoupling}

Since we do not start with a Lagrangian but the equations of motion,
the existence of the negative component of the generators
does not immediately mean the existence of the negative norm states
(\textit{i.e.}, fields with wrong sign kinetic terms).
However, we aim to construct effective actions with respect to various
values of $C^\mu_A$ and it is plausible that we can introduce a
prescription to deal with the ghost fields.
In this paper, we take the strategy used in
\cite{Ho:2008ei,Bandres:2008kj,Gomis:2008be}, where the shift symmetry
existing for the center components are gauged and these components
are gauged away.
At the same time, the equations of motion for newly introduced gauge
fields provide constraints for the paired components, and they become
non-dynamical.

As we will see, $X^I$ and $\Psi$ have an ordinary Lagrangian
description even for the non-Abelian case.
For the self-dual three-form field $H_{\mu\nu\rho}$, the treatment is
slightly different for each situation.
We will give a sketch of the decoupling mechanism  here and will
supplement comments later in each subsection.

Since the interaction terms always involve the structure constant
$f^{BCD}{}_A$, $X^I_{\ul{0}}$, $X^I_{\ul{a}}$,
$\Psi_{\ul{0}}$ and $\Psi_{\ul{a}}$ components
appear in the action only through the kinetic terms:
\begin{align}
    \label{eq:ghost_action}
    \mathcal{L}_{gh} =&
-(D_\mu X^I)_0 (D^\mu X^I)_{\ul{0}}
-(D_\mu X^I)_a (D^\mu X^I)_{\ul{a}}
\nn\\&
+\frac{i}{2} \left(
\bar\Psi_0 \Gamma^\mu D_\mu \Psi_{\ul{0}}
+\bar\Psi_a \Gamma^\mu D_\mu \Psi_{\ul{a}}
\right)
\nn\\=&
-(D_\mu X^I)_\alpha (D^\mu X^I)_{\ul{\alpha}}
+\frac{i}{2}
\bar\Psi_\alpha \Gamma^\mu D_\mu \Psi_{\ul{\alpha}} \,,
\end{align}
where $\alpha=(0,a)$ and $a=1,\dots,d$.
It should be noted that this kinetic term correctly reproduces the
kinetic term of
the $u^{\ul{\alpha}}$ part.
Because of the restricted form of the gauge field $\tilde{A}_\mu$,
the covariant derivative has to take the following form:
\begin{align}
    \left( D_\mu \phi \right)_{\ul{\alpha}}
= \partial_\mu \phi_{\ul{\alpha}} + (\text{terms not including
  $\phi_{\ul{\alpha}}$}) \,,
\end{align}
where $\phi_{\ul{\alpha}}$ means $X^I_{\ul{\alpha}}$ or $\Psi_{\ul{\alpha}}$.
Therefore, there exist the following shift symmetries:
\begin{align}
    X^I_{\ul{\alpha}} \rightarrow X^I_{\ul{\alpha}} +
    \xi^I_{\ul{\alpha}} \,,
\qquad
     \Psi_{\ul{\alpha}} \rightarrow \Psi_{\ul{\alpha}} +
    \eta_{\ul{\alpha}} \,,
\end{align}
with constant $\xi^I_{\ul{\alpha}}$ and $\eta_{\ul{\alpha}}$.
We now promote these shift symmetries to gauged ones (space-time dependent):
\begin{align}
    \xi^I_{\ul{\alpha}} \rightarrow \xi^I_{\ul{\alpha}}(x) \,,
\qquad
    \eta_{\ul{\alpha}} \rightarrow \eta_{\ul{\alpha}}(x) \,, 
\end{align}
by introducing gauge fields:
\begin{align}
    \partial_\mu X^I_{\ul{\alpha}} \rightarrow 
\partial_\mu X^I_{\ul{\alpha}} +    a^I_{\mu\; \ul{\alpha}} \,,
\qquad
    \partial_\mu \Psi_{\ul{\alpha}} \rightarrow 
\partial_\mu \Psi_{\ul{\alpha}} +    b_{\mu\; \ul{\alpha}} \,,
\end{align}
that obey the transformation law:
\begin{align}
    a^I_{\mu\; \ul{\alpha}} \rightarrow 
a^I_{\mu\; \ul{\alpha}} - \partial_\mu \xi^I_{\ul{\alpha}}(x) \,,
\qquad
    b_{\mu\; \ul{\alpha}} \rightarrow 
b_{\mu\; \ul{\alpha}} - \partial_\mu \eta_{\ul{\alpha}}(x) \,.
\end{align}
Now we can gauge away $X^I_{\ul{\alpha}}$ and $\Psi_{\ul{\alpha}}$
by a gauge choice, and the equations of motion of these gauge
fields impose the constraints:
\begin{align}
    \partial_\mu X^I_\alpha = \Psi_\alpha =0 \,,
\end{align}
on the conjugate fields $X^I_\alpha$ and $\Psi_\alpha$,
which used to satisfy the free equations of motion:
\begin{align}
\partial^\mu    \partial_\mu X^I_\alpha = 
\Gamma^\mu \partial_\mu \Psi_\alpha =0 \,.
\end{align}
In this way, we can eliminate the unwanted fields, and instead 
obtain some
``moduli'' fields.
In the following analysis of the equations of motion, we will assume
that these ghost fields have already been eliminated and
\begin{align}
    X^I_0 = \lambda^I_0 \,,
\quad
 X^I_a = \lambda^I_a \,,
\quad
\Psi_0=\Psi_a =0 \,, \label{Ghost-decouple-X-psi}
\end{align}
are imposed, where $\lambda^I_0$ and $\lambda_a^I$
are certain constant vectors and will be identified as moduli of the theory.

For the three-form field $H_{\mu\nu\rho}$, the situation is more
complicated since it does not allow a simple Lagrangian description due
to 
 self-duality.
However, we can still observe 
a shift symmetry:
\begin{align}
  H_{\mu\nu\rho\; \ul{\alpha}} \rightarrow   
H_{\mu\nu\rho\; \ul{\alpha}} + \zeta_{\mu\nu\rho\; \ul{\alpha}} \,,
\end{align}
in the equations of motion for $H_{\mu\nu\rho\; \ul{\alpha}}$,
and then by gauging it we can eliminate $H_{\mu\nu\rho\; \ul{\alpha}}$.
To do so, we introduce a gauge field 
that transforms as
\begin{align}
 G_{\mu\nu\rho\; \ul{\alpha}} \rightarrow G_{\mu\nu\rho\; \ul{\alpha}}
- \zeta_{\nu\rho\sigma\; \ul{\alpha}} \,.
\end{align}
If the kinetic term of $H$ field were like:
\begin{align}
  \frac{1}{2}\frac{1}{3!} H^{\mu\nu\rho}_\alpha 
\left( H_{\mu\nu\rho \; \ul{\alpha}} + G_{\mu\nu\rho\; \ul{\alpha}} \right) \,,
\end{align}
the gauge field equation of motion would lead to the constraints:
\begin{align}
  H_{\mu\nu\rho\; 0}= H_{\mu\nu\rho\; a}=0 \,. \label{Ghost-decouple-H}
\end{align}
However, because of 
self-duality,
it does not go easily.
We will discuss the treatment of $H_{\mu\nu\rho\; 0}$ and
$H_{\mu\nu\rho \; a}$ in each case separately.

\subsection{Generic case}
\label{sec:gen_case}

\paragraph{$C$ fields and gauge fields}

In this subsection, we consider the case with all $C_A^\mu$ being
nonzero, $C^\mu = C_{0}^\mu u^0 + C_a^\mu u^a + C_{(i\vec{m})}^\mu
T^{(i\vec{m})}$.
This is the case 1 (including the case 2) of the classification
  in
section \ref{sec:constraints}.
As seen in section \ref{sec:constraints}, $C_a^\mu$ and
$C_{(i\vec{m})}^\mu$ components are proportional to $C_0^\mu$
component:
\begin{align}
    C_a^\mu = v_a C^\mu_0 \,,
\qquad
C_{(i\vec{m})}^\mu = v_{(i\vec{m})} C_0^\mu \,.
\end{align}
It should be  noted that $C_0^\mu$ are constant due to the equation of
motion.
Thus it is always possible, by Lorentz rotation,
to align $C_0^\mu$ into one direction, 
say $\mu=\tilde\mu$,
and to make the other components vanish, $C_0^{\mu \neq \tilde\mu}=0$.
Therefore, in this case, we assume that
the auxiliary field $C_0^\mu$ takes nonzero
value only for $\mu=\tilde\mu$ without loss of generality.
We also assume that $C_{\ul{0}}^\mu$ and $C_{\ul{a}}^\mu$ components
are not coupled to the other physical fields,
$\partial_\nu C_{\ul{0}}^\mu =\partial_\nu  C_{\ul{a}}^\mu=0$.
This condition, together with (\ref{eq:DC_ghost1}), implies
\begin{align}
    v_a \AAB{\mu}{a}{\ul{0}} =& 
-v_{(i\vec{m})} \AAB{\mu}{(i\vec{m})}{\ul{0}} \,,
\qquad 
\AAB{\mu}{0}{\ul{a}} =
- v_{(i\vec{m})} \AAB{\mu}{(i\vec{m})}{\ul{a}} \,,
\end{align}
for non-zero $C_0^\mu$.
The anti-symmetry $\AAB{\mu}{0}{\ul{a}}=-\AAB{\mu}{a}{\ul{0}}$
gives the relation
\begin{align}
v_{(i,-\vec{m})}
\left( 
 \AAB{\mu}{0}{(i\vec{m})}
+ v_a   \AAB{\mu}{a}{(i\vec{m})}
\right)=0 \,.
\end{align}
Now we 
consider the constant values of the $C_{0,a}$ fields, $C_0^{\tilde\mu}$
and $v_a$, as moduli of the effective theory.
Then $v_{(i\vec{m})}$ is also preferred to be taken as an unrestricted
parameter here.
This requires the condition
$ \AAB{\mu}{0}{(i\vec{m})}  =- v_a   \AAB{\mu}{a}{(i\vec{m})}$.
Note that though this $v_{(i\vec{m})}$ turns out to be zero
as a result of the gauge
field equation of motion, 
we leave $v_{(i\vec{m})}$ unrestricted for a while.
On such $v_{(i\vec{m})}$, the condition $(\tilde{D}_{\mu}v)_{(i\vec{m})}=0$ is imposed due to (\ref{Cim constraint}).

By employing the relationship $\AAB{\mu}{B}{A} = A_{\mu
  CD}f^{CDB}{}_{A}$ and $\AAB{\mu}{a}{(i\vec{m})}=  im^aA_{\mu
  0(i\vec{m})}$,
together with the above relationships such as
$\AAB{\mu}{0}{(i\vec{m})}= -v_a\AAB{\mu}{a}{(i\vec{m})}$,
the gauge fields are represented by the ones without tilde as
\begin{align}
\label{eqs:relation-A1}
    \AAB{\mu}{a}{(i\vec{m})} =&
i m^a A_{\mu (i\vec{m})}
\,,
\\    
    \AAB{\mu}{0}{(i\vec{m})} =& -i m^a v_a A_{\mu (i\vec{m})} \,,
\\
\AAB{\mu}{0}{\ul{a}} =&  i m^a v_{(i,-\vec{m})} A_{\mu(i\vec{m})} \,,
\\
    \AAB{\mu}{(i\vec{m})}{(j\vec{n})} =&
\fijk{k}{i}{j} A_{\mu (k,\vec{n}-\vec{m})} \,,
\\
\label{eqs:relation-A5}
    \AAB{\mu}{(i\vec{m})}{(i\vec{m})} =&
i m^a a_{\mu \; a} \,,
\end{align}
where we have defined
$A_{\mu \; 0(i\vec{m})} \equiv A_{\mu (i\vec{m})}$
and $A_{\mu \; a0} \equiv a_{\mu a}$.
So the gauge fields are all 
represented by these two gauge fields.
From (\ref{tilA_ntilA1}), this relation implies
\begin{align}
    \fijk{j}{k}{i} A_{\mu \, (j\vec{n})(k,\vec{m}-\vec{n})} =0 \,,
\qquad
 A_{\mu \, a (i\vec{m})} = v_a A_{\mu (i\vec{m})} \,.
\end{align}

From $\left( D_{\mu}C^{\rho} \right)_A = 0$,
 there are also the
constraints 
on
the field strength $\tilde{F}_{\mu\nu}$ as
\begin{equation}
0 = ([D_{\mu}, D_{\nu}]C^{\rho} )_{A} = 
\FAB{\mu}{\nu}{B}{A}C^{\rho}_{B}, \label{eq:Costraint-F}
\end{equation}
for any gauge indices $A$.
For each gauge index, the equation (\ref{eq:Costraint-F}) is 
written as
\begin{align}
\label{eq:F-equations1}
0 =&
 \left( \FAB{\mu}{\nu}{0}{\ul{a}}
+v_{(i\vec{m})}\FAB{\mu}{\nu}{(i\vec{m})}{\ul{a}}\right)C^{\rho}_{0} 
\,,\\
\label{eq:F-equations2}
0 =&
 v_{(i\vec{m})}\left(\FAB{\mu}{\nu}{(i\vec{m})}{\ul{0}}
+ v_a\FAB{\mu}{\nu}{(i\vec{m})}{\ul{a}}\right)C^{\rho}_{0} 
\,,\\
0 =&
(\FAB{\mu}{\nu}{(i\vec{m})}{(i\vec{m})}v_{(i\vec{m})} 
+ \FAB{\mu}{\nu}{(j\vec{n})}{(i\vec{m})}v_{(j\vec{n})}) C^{\rho}_{0} \, .
\label{eq:F-equations3}
\end{align}
Since there is one non-vanishing $C_0^{\bar \mu}$,
we have the relations 
between the field strengths:
\begin{gather}
  \FAB{\mu}{\nu}{{a}}{0} = -v_{(i,-\vec{m})}
  \FAB{\mu}{\nu}{a}{(i\vec{m})} \,,
\quad
  \FAB{\mu}{\nu}{0}{(i\vec{m})} =
-v_a  \FAB{\mu}{\nu}{a}{(i\vec{m})} \,,
\nn \\
v_{(i\vec{m})} \FAB{\mu}{\nu}{(i\vec{m})}{(i\vec{m})}
= -v_{(j\vec{n})} \FAB{\mu}{\nu}{(j\vec{n})}{(i\vec{m})} \,.
  \label{eqs:relation-F}
\end{gather}
The 
diagonal part can be written in terms of only
$a_{\mu \; a}$,
\begin{equation}
\FAB{\mu}{\nu}{(i\vec{m})}{(i\vec{m})}=
-i m^a f_{\mu\nu\; a} \,,
\qquad
f_{\mu\nu\, a} \equiv \partial_{\mu}a_{\nu\; a}
-\partial_{\nu}a_{\mu\; a} \,.
\end{equation}
It should be noted that
for the second equation of (\ref{eqs:relation-F}),
we use the fact that
$v_{(i\vec{m})}$ is unconstrained,
but by using the explicit form of the gauge fields
(\ref{eqs:relation-A1})--(\ref{eqs:relation-A5})
this can also be confirmed.

Analogously,
the following identity:
\begin{equation}
\label{eq:DF-impose}
0 = \left(D^{\mu}[D_{\mu}, D_{\nu}]C^{\rho} \right)_{A} = 
  \left(D^\mu \tilde{F}_{\mu\nu} \right)^B{}_A C^{\rho}_{B} \,,
\end{equation}
provides the same relations among the covariant derivatives of the
field strength:
\begin{gather}
  \left(D^\mu \tilde{F}_{\mu\nu} \right)^{0}{}_{\ul{a}}
= v_{(i,-\vec{m})}
\left(D^\mu \tilde{F}_{\mu\nu} \right)^{a}{}_{(i\vec{m})}
\,,
\quad
\left(D^\mu \tilde{F}_{\mu\nu} \right)^{0}{}_{(i\vec{m})}
= -v_a
\left(D^\mu \tilde{F}_{\mu\nu} \right)^{a}{}_{(i\vec{m})}
\,,
  \label{eqs:relation-DF1}
 \\
v_{(i\vec{m})}
\left(D^\mu \tilde{F}_{\mu\nu} \right)^{(i\vec{m})}{}_{(i\vec{m})}
= -v_{(j\vec{n})}
\left(D^\mu \tilde{F}_{\mu\nu} \right)^{(j\vec{n})}{}_{(i\vec{m})}
\,,
  \label{eqs:relation-DF2}
\end{gather}
and we will use these relations later to simplify the gauge field
equations of motion.
Again, the diagonal part becomes
\begin{align}
      \left(  D^{\mu}\tilde{F}_{\mu\nu}\right)^{(i\vec{m})}{}_{(i\vec{m})} 
=&
-i m^a \partial^{\mu}f_{\mu\nu \; a } \,.
\label{eq:DF_abelian}
\end{align}

Next, we consider the equations (\ref{feq6}):
\begin{align}
  \thcommt{C^\mu}{D_\mu \phi}{T^B}_A = 0 \,,
\end{align}
where $\phi$ denotes any of $X^I$, $\Psi$ or $H_{\mu\nu\rho}$.
From (\ref{feq1})--(\ref{feq3}),
(\ref{Ghost-decouple-X-psi}) and (\ref{Ghost-decouple-H}),
it is easy to see that
\begin{align}
\left(  D_\mu \phi\right)_0 
= \left(  D_\mu \phi\right)_a 
=0 \,,
\end{align}
and
thus $u^0$ and $u^a$ components do not enter this constraint
equation.
$(D_\mu \phi)_{\ul{0}}$ and $(D_\mu \phi)_{\ul{a}}$ are also missing
since they are in the center.
Therefore, these equations involve only in $(D_\mu \phi)_{(i\vec{m})}$ components.
The independent equations from (\ref{feq6}) turn out to be
\begin{align}
\label{feq6-1}
  m^a C_0^\mu \left(D_\mu \phi \right)_{(i\vec{m})} =& 0 \,,
\\
\label{feq6-2}
f^{ij}{}_k C_0^\mu \left(D_\mu \phi \right)_{(i\vec{m})} =& 0 \,,
\end{align}
where in the first equation (\ref{feq6-1}) no summation with respect to
 $\vec{m}$ is taken.
Thus, for non-zero mode $\vec{m} \neq \vec{0}$, the first equation
(\ref{feq6-1}) means 
$C_0^\mu \left(D_\mu \phi \right)_{(i\vec{m})} =0$, which also solves
the second equation (\ref{feq6-2}).
For zero mode $\vec{m}=\vec{0}$, the first equation is trivial.
If the index $i$ for $\left(D_\mu \phi \right)_{(i\vec{0})}$ satisfies
$f^{ij}{}_k=0$ for all $j$ and $k$, namely it belongs to 
an Abelian
sub-algebra, the second equation is satisfied.
Otherwise, the second one again gives the restriction $C_0^\mu
\left(D_\mu \phi \right)_{(i\vec{0})} =0$.
Recall that $C_0^\mu$ is nonzero only for $\mu=\tilde\mu$.
We then summarize the result;
\begin{itemize}
      \item For non zero modes, $\left(D_\mu \phi \right)_{(i,\vec{m}\neq \vec{0})}$,
and the zero mode with indices that are not in 
any Abelian sub-algebra,
$\left(D_\mu \phi \right)_{(i\vec{0})}$ with $f^{ij}{}_k \neq 0$ for
some $j$ and $k$, the constraint imposes the condition:
\begin{align}
\label{eq:dim_rec_byC}
    C_0^{\tilde\mu} \left(D_{\tilde\mu} \phi \right)_{(i\vec{m})} =0 \,,
\end{align}
that is, the covariant derivatives for the fields in tensor multiplets
are suppressed in the direction of $C_0^{\tilde\mu} \neq 0$.
We can therefore see that the world volume is dimensionally reduced in this
$\tilde\mu$ direction.
  \item For zero modes associated with Abelian sub-algebra, no reduction
occurs.
However, 
these modes (i.e., zero modes associated with Abelian sub-algebras) will
turn out to be decoupled from the other modes that we are interested
in, and therefore will not be included in the effective Lagrangians
we will discuss.
\end{itemize}
With these structures in mind, we next analyze the equations of $X^I$
and $\Psi$.

\paragraph{$X^I$ and $\Psi$ part}

It is not difficult to see that the 
equations of
motion (\ref{feq1}) and (\ref{feq3}) can be obtained by the following
Lagrangians:
\begin{align}
    \mathcal{L}_X =&
-\frac{1}{2} \left\langle
 \left(D_\mu X^I \right) ,  \left(D^\mu X^I \right) \right\rangle
+\frac{1}{4}  \left\langle
              \thcommt{C^\mu}{X^I}{X^J} ,
             \thcommt{C_\mu}{X^I}{X^J}   \right\rangle
\nn\\=&
-\frac{1}{2} \left(D_\mu X^I \right)_{(i,-\vec{m})} \left(D^\mu X^I \right)_{(i\vec{m})}
+\frac{1}{4} \thcommt{C^\mu}{X^I}{X^J}_{(i\vec{m})}
             \thcommt{C_\mu}{X^I}{X^J}_{(i,-\vec{m})}
\,, 
\label{eq:Lag_X_gen}
\\
\mathcal{L}_\Psi =&
\frac{i}{2}  \left\langle
    \Bar\Psi , \Gamma^\mu  \left( D_\mu \Psi \right)   \right\rangle
+\frac{1}{2}
\left\langle \Bar\Psi , 
\Gamma_\nu \Gamma^I  \thcommt{C^\nu}{X^I}{\Psi} \right\rangle
\nn\\=&
\frac{i}{2} \Bar\Psi_{(i,-\vec{m})} \Gamma^\mu \left( D_\mu \Psi \right)_{(i\vec{m})}
+\frac{1}{2} \Bar\Psi_{(i,-\vec{m})} \Gamma_\nu \Gamma^I 
  \thcommt{C^\nu}{X^I}{\Psi}_{(i\vec{m})} \,,
\label{eq:Lag_Psi_gen}
\end{align}
where in each second line we have eliminated the ghost fields by using
the shift symmetry.

First, we look at the covariant derivatives.
By using
\begin{align}
    \AAB{\mu}{0}{(i\vec{m})} =& - v_a \AAB{\mu}{a}{(i\vec{m})} \,,
\quad
X_0^I = \lambda_0^I \,,
\quad
X_a^I = \lambda_a^I \,,
\end{align}
we have
\begin{align}
    \left( D_\mu X^I \right)_{(i\vec{m})} =&
\partial_\mu X^I_{(i\vec{m})}
- \AAB{\mu}{(j\vec{n})}{(i\vec{m})} X^I_{(j\vec{n})}
- \AAB{\mu}{(i\vec{m})}{(i\vec{m})} X^I_{(i\vec{m})}
- \AAB{\mu}{0}{(i\vec{m})} X^I_{0}
- \AAB{\mu}{a}{(i\vec{m})} X^I_{a}
\nn\\=&
\left( \tilde{D}_\mu X^I \right)_{(i\vec{m})}
- \tau_a^I \left( \partial^a A_{\mu} \right)_{(i\vec{m})} \,,
\end{align}
where we have defined $\tau_a^I \equiv \lambda_a^I - v_a \lambda_0^I$
and used (\ref{eqs:relation-A1})--(\ref{eqs:relation-A5}).
Here the covariant derivative $\tilde{D}_{\mu}$
is defined as 
(\ref{eq:def_Tilde_D}) in Appendix \ref{sec:assum_gauge}.
$m^a$ can be regarded as the momentum associated with the internal
direction $y_a$, and we replace it with the derivative with respect to 
$y_a$, 
$i m^a \phi_{(i\vec{m})} = (\partial^a \phi)_{(i\vec{m})}$.
We will often employ this identification later.
We now decompose the bosonic field $X^I_{(i\vec{m})}$ into the
components that are parallel to $\tau_a^I$ and the ones perpendicular
to that, by following the idea of \cite{Kobo:2009gz}.
We first introduce a projector:
\begin{align}
    \label{eq:def_PIJ}
P^I_J \equiv & \delta^I_J - \tau_a^I \pi^a_J \,,
\qquad
\pi^a_I \tau_b^I = \delta^a_b \,,
\end{align}
where the conjugate vectors $\pi^a_I$ are defined by
\begin{align}
    \pi^a_I =& \delta_{IJ} g^{ab} \tau_b^J \,,
\end{align}
where $g_{ab} = \tau_a^I \tau_b^I$ is a metric constructed from
``vielbein'' $\tau_a^I$ and is assumed to be invertible.
By using this projector, we define
\begin{align}
    X_{(i\vec{m})}^I = P^I_J X_{(i\vec{m})}^J + \tau_a^I
    Y^a_{(i\vec{m})} \,,
\end{align}
where $Y^a_{(i\vec{m})} \equiv \pi^a_J X^J_{(i\vec{m})}$.
Note that $\tau_a^I$ and $\pi^a_I$ are constant and also covariantly
constant with respect to $\tilde{D}_\mu$ since the gauge rotation in
$\tilde{D}_\mu$ is only involved in $(i\vec{m})$ index.
This fact 
leads to
\begin{align}
    \left( D_\mu X^I \right)_{(i\vec{m})} =&
P^I_J \left( \tilde{D}_\mu X^J \right)_{(i\vec{m})}
+ \tau_a^I \left( \tilde{D}_\mu Y^a - \partial^a A_\mu
\right)_{(i\vec{m})} \,.
\end{align}
Since $\Psi_0=\Psi_a=0$, the fermion field kinetic term simply becomes
\begin{align}
    \left( D_\mu \Psi \right)_{(i\vec{m})} =&
 \left( \tilde{D}_\mu \Psi \right)_{(i\vec{m})} \,.
\end{align}
Now we move on to the potential terms.
By using the same moduli fields, we have
\begin{align}
    \label{eq:CXX_term1}
&
\thcommt{C^\mu}{X^I}{X^J}_{(i\vec{m})}
\nn\\ =&
C_0^\mu\left(
m^a \tau_a^I \left( X^J - \lambda_0^J v \right)
-m^a \tau_a^J \left( X^I - \lambda_0^I v \right)
+\commt{X^I - \lambda_0^I v}{X^J - \lambda_0^J v}
\right)_{(i\vec{m})}
\nn\\=&
C_0^\mu
\left(
\commt{(P\tilde{X})^I}{(P\tilde{X})^J}
-i \tau_a^I \hat\nabla^a (P\tilde{X})^J
+i \tau_a^J \hat\nabla^a (P\tilde{X})^I
- i \tau_a^I \tau_b^J f^{ab}
\right)_{(i\vec{m})} \,,
\end{align}
where $\tilde{X}^I_{(i\vec{m})}=X^I_{(i\vec{m})} - \lambda_0^I
v_{(i\vec{m})}$
is a shifted $X^I_{(i\vec{m})}$ and we
decompose it into $\tilde{X}^I_{(i\vec{m})} =
 P^I_J \tilde{X}^J_{(i\vec{m})}
+ \tau_a^I \tilde{Y}^a$
with $\tilde{Y}^a_{(i\vec{m})} 
= Y^a_{(i\vec{m})} - \pi^a_J \lambda_0^J v_{(i\vec{m})}$.
A new covariant derivative $\hat\nabla^a$ is defined by
\begin{align}
    \left( \hat\nabla^a \phi \right)_{(i\vec{m})} =&
\left( \partial^a \phi + i
\commt{\tilde{Y}^a}{\phi} \right)_{(i\vec{m})} \,,
\end{align}
namely, $\tilde{Y}^a_{(i\vec{m})}$ is treated as a new gauge field,
and $m^a$ is transformed into the derivative in $a$ direction, $\partial^a = im^a$.
$f^{ab}_{(i\vec{m})}$ is the field strength corresponding to this
gauge field:
\begin{align}
  f^{ab}_{(i\vec{m})} =&
\left( \partial^a \tilde{Y}^b
- \partial^b \tilde{Y}^a
+ i \commt{\tilde{Y}^a}{\tilde{Y}^b} \right)_{(i\vec{m})} \,.
\end{align}
In the fermionic part, we have the potential term
including
\begin{align}
  \thcommt{C^\nu}{X^I}{\Psi}_{(i\vec{m})} =&
C_0^\nu m^a \tau_a^I \Psi_{(i\vec{m})}
+C_0^\nu \commt{X^I- \lambda_0^I v}{\Psi}_{(i\vec{m})}
\nn\\=&
-i C_0^\nu \tau_a^I \left(\hat\nabla^a \Psi \right)_{(i\vec{m})}
+C_0^\nu P^I_J \commt{\tilde{X}^J}{\Psi}_{(i\vec{m})}
 \,.
\end{align}
Collecting all the results we can spell out the effective Lagrangian
for $X^I$ and $\Psi$ fields.
Note that since $\tilde{D}_\mu v_{(i\vec{m})}=0$, we can replace
$(\tilde{D} X^I)_{(i\vec{m})}$ with $(\tilde{D}
\tilde{X}^I)_{(i\vec{m})}$, the same for $Y$, and then remove all
tilde from $\tilde{X}^I$ and $\tilde{Y}$
by redefining the fields.
Therefore, $v_{(i\vec{m})}$ plays a trivial role in $X^I$ and $\Psi$
part Lagrangian.
Later, we will see that the gauge field equations of motion force
$v_{(i\vec{m})}=0$, but this change does not affect this part of the
Lagrangian.
Finally the effective Lagrangian of this part becomes
\begin{align}
  \label{eq:gen_C_XPsi_Lag}
  \mathcal{L}_{X+\Psi}=&
-\frac{1}{2} P_{IJ}
   \left( \tilde{D}^\mu X^I \right)_{(i,-\vec{m})}
   \left( \tilde{D}_\mu X^J \right)_{(i\vec{m})}
-\frac{1}{2} C^2 P_{IJ} g_{ab} 
   \left( \hat{\nabla}^a X^I \right)_{(i,-\vec{m})}
   \left( \hat{\nabla}^b X^J \right)_{(i\vec{m})}
\nn\\ &
+\frac{1}{4}C^2 P_{IK} P_{JL} 
  \commt{X^I}{X^J}_{(i,-\vec{m})}
  \commt{X^K}{X^L}_{(i\vec{m})}
-\frac{1}{4} C^2 g_{ac} g_{bd} 
  f^{ab}_{(i,-\vec{m})} f^{cd}_{(i\vec{m})}
\nn\\ &
-\frac{1}{2} g_{ab}
 \left(\tilde{D}_\mu Y^a - \partial^a A_\mu \right)_{(i,-\vec{m})}
 \left(\tilde{D}^\mu Y^b - \partial^b A^\mu \right)_{(i\vec{m})}
\nn\\ &
+\frac{i}{2} \Bar\Psi_{(i,-\vec{m})}
 \left( \Gamma^\mu ( \tilde{D}_\mu \Psi )
-C_0^\nu \Gamma_\nu \Gamma^I \tau_a^I ( \hat{\nabla}^a \Psi )
 \right)_{(i\vec{m})} 
\nn\\ &
+\frac{1}{2} \Bar\Psi_{(i,-\vec{m})}
 C_0^\nu \Gamma_\nu 
 \Gamma^I  P^I_J \commt{X^J}{\Psi}_{(i\vec{m})}
 \,.
\end{align}
In the fermion kinetic term,  $\Gamma^\mu \partial_{\mu}
-C_0^\nu \Gamma_\nu \Gamma^I \tau_a^I\partial^a$ is a Dirac operator in the sense of 
\begin{eqnarray}
(i\Gamma^\mu \partial_{\mu}
+C_0^\mu \Gamma_\mu \Gamma^I \tau_a^Im^a)
(-i\Gamma^\nu \partial_{\nu}
+C_0^\nu \Gamma_\nu \Gamma^J \tau_b^Jm^b)
=\partial^{\mu}\partial_{\mu}-C_0^2g_{ab}m^am^b.
\end{eqnarray}
Because  the formulation of \cite{Lambert:2010wm} that we use does not have the  $SO(1,10)$ covariance, it gives different structures of Gamma matrix between  $\tilde{D}_{\mu}$ and $\hat{\nabla}^a $.

\paragraph{Gauge field $H_{\mu\nu\rho}$ and $\tilde{F}_{\mu\nu}$}

Finally, we consider the self-dual three-form field $H_{\mu\nu\rho\; A}$
and the field strength $\FAB{\mu}{\nu}{B}{A}$.
Because of the constraint (\ref{feq4}), which is in the bracket form:
\begin{align}
\label{feq4-2}
  \FAB{\mu}{\nu}{B}{A} =&
i \thcommt{C^\rho}{H_{\mu\nu\rho}}{T^B}_A \,,
\end{align}
$\tilde{F}_{\mu\nu}$ and $H_{\mu\nu\rho}$ are not independent degrees
of freedom when $C^\rho \neq 0$.
Therefore we can rewrite $H_{\mu\nu\rho}$ equation of motion
(\ref{feq2}) in terms of $\tilde{F}_{\mu\nu}$.
Let us take a close look at the equation of motion for
$H_{\mu\nu\rho\; A}$ (\ref{feq2}) that takes the form of the
Bianchi identity.
Thanks to the self-duality condition, $H_{\mu\nu\rho\;
  A}=\frac{1}{3!}\epsilon_{\mu\nu\rho\sigma\tau\lambda}H^{\sigma\tau\lambda}{}_A$,
the first derivative terms can be changed by multiplying the epsilon
tensor\footnote{See the appendix for our convention for the
  antisymetrization.}:
\begin{align}
  \epsilon_{\mu\nu\rho\lambda\sigma\tau}
D^{[\mu} H^{\nu\rho\lambda]}{}_A =&
3! D^\mu H_{\mu\sigma\tau\; A} \,,
\end{align}
and then (\ref{feq2}) can be written as the usual equation of motion:
\begin{align}
  D^\mu H_{\mu\nu\lambda\; A}
+2i \thcommt{C_{[\nu}}{X^I}{D_{\lambda]} X^I}_A
- \thcommt{C_{[\nu}}{\Bar\Psi}{\Gamma_{\lambda]} \Psi}_A
=0 \,.
\end{align}
We may define a two-form ``current'' $J_{\nu\lambda\; A}$ as
\begin{align}
\label{eq:Def_J}
  J_{\nu\lambda\; A} =&
2i \thcommt{C_{[\nu}}{X^I}{D_{\lambda]} X^I}_A
- \thcommt{C_{[\nu}}{\Bar\Psi}{\Gamma_{\lambda]} \Psi}_A \,,
\end{align}
and the equation of motion is now
\begin{align}
  D^\mu H_{\mu\nu\lambda\; A} =& -J_{\nu\lambda\; A} \,.
\end{align}
It should be noted that because of the three-bracket, 
$J_{\mu\nu\; 0}=J_{\mu\nu\; a}=0$ immediately follows.
For $A=\ul{0}, \ul{a}$ case, we have
\begin{align}
  \partial^\mu H_{\mu\nu\lambda\; \ul{0}}
-\tilde{A}^{\mu\; B}{}_{\ul{0}} H_{\mu\nu\lambda\; B}
 =& -J_{\nu\lambda\; \ul{0}}
\end{align}
and the same for $\ul{a}$.
$H_{\mu\nu\lambda\; \ul{0}}$ appears only here among the equations of
motion.
Once we remove this ghost mode by gauging the shift symmetry, this
equation becomes a constraint for $J_{\nu\lambda\; \ul{0}}$.
As we will see, however, $J_{\nu\lambda\; \ul{0}}$ will decouple from
the rest of the dynamics, and then we can safely assume that this
constraint is always satisfied.

Now we multiply $D^{\mu}$ to (\ref{feq4-2}).
Since
\begin{align}
  \left(D^\mu \thcommt{C^\rho}{H_{\mu\nu\rho}}{T^B} \right)_A
=&
\thcommt{D^\mu C^\rho}{H_{\mu\nu\rho}}{T^B}_A
+\thcommt{C^\rho}{D^\mu H_{\mu\nu\rho}}{T^B}_A \, \nn \\
=&
\thcommt{C^\rho}{D^\mu H_{\mu\nu\rho}}{T^B}_A \, ,
\end{align}
we have
\begin{align}
  \left( D^\mu \tilde{F}_{\mu\nu} \right)^B{}_A
=& i \thcommt{C^\lambda}{D^\mu H_{\mu\nu\lambda}}{T^B}_A 
\nn\\ =&
-i \thcommt{C^\lambda}{J_{\nu\lambda}}{T^B}_A 
\nn\\=&
C_C^\lambda J_{\nu\lambda\; D} f^{CDB}{}_A \,.
\label{eq:DF=J_rel}
\end{align}
Because $J_{\nu\lambda\; \ul{0}}$ and $J_{\nu\lambda\; \ul{a}}$ are in
the center, they do not contribute to this equation as we have anticipated.
The independent equations are
\begin{align}
\label{eq:DF-current1}
  \left( D^\mu \tilde{F}_{\mu\nu} \right)^0{}_{\ul{a}} =&
i m^a v_{(i,-\vec{m})} C_0^\lambda  J_{\nu\lambda\; (i\vec{m})}   \,,
\\
\label{eq:DF-current2}
  \left( D^\mu \tilde{F}_{\mu\nu} \right)^0{}_{(i\vec{m})} =&
-i m^a v_a C_0^\lambda  J_{\nu\lambda\; (i\vec{m})} 
-i[v, C^{\lambda}_{0}J_{\nu\lambda}]_{(i\vec{m})}
\,,
\\
\label{eq:DF-current3}
  \left( D^\mu \tilde{F}_{\mu\nu} \right)^a{}_{(i\vec{m})} =&
i m^a C_0^\lambda J_{\nu\lambda\; (i\vec{m})}   \,,
\\
\label{eq:DF-current4}
  \left( D^\mu \tilde{F}_{\mu\nu} \right)^{(i\vec{m})}{}_{(i\vec{m})} =&
0 \,,
\\
\label{eq:DF-current5}
  \left( D^\mu \tilde{F}_{\mu\nu} \right)^{(i\vec{m})}{}_{(j\vec{n})} =&
 f^{ki}{}_j
C_0^\lambda J_{\nu\lambda\; (k,\vec{n}-\vec{m})}   \,.
\end{align}
Therefore now $C_0^\lambda J_{\nu\lambda\; (i\vec{m})}$ plays the role
of the source current for $\tilde{F}_{\mu\nu}$.
Comparing (\ref{eq:DF-current1})-(\ref{eq:DF-current5})
with (\ref{eqs:relation-DF1}) and (\ref{eq:DF_abelian}),
we have
\begin{align}
0 =& m^a \partial^{\mu} f_{\mu\nu \;a } , \, \label{eq:background}\\
0 =& [v, C^{\lambda}_{0}J_{\nu\lambda}]_{(i\vec{m})}.
\label{eq:vim zero}
\end{align}
Let us examine (\ref{eq:background}) first.
This relation holds for arbitrary $\vec{m}$, and then it means
$\partial^\mu f_{\mu\nu \; a} =0$.
If we differentiate the effective Lagrangian 
(\ref{eq:gen_C_XPsi_Lag}) with respect to $a_{\mu \; a}$,
there appears a non-zero current composed by
$X^I$ and $\Psi$ for this equation.
So the gauge fields $a_{\mu\; a}$ must be regarded as the background ones,
of which we do not consider the variation.
Because the background fields $a_{\mu\; a}$ satisfy the 
vacuum equation of motion,
we will assume the simplest solution $a_{\mu \; a}=0$ here
\footnote{The background fields $a_{\mu\; a}$ can be
represented by 2-form gauge fields $b_{\mu\nu\; a}$ 
and $b_{\mu\nu\; 0}$.
This is discussed in the appendix \ref{sec:amu and 2-form}
in detail.}.
The second equation (\ref{eq:vim zero}) means that $v_{(i\vec{m})}$
must be zero apart from the zero mode $v_{(i\vec{0})}$\footnote{%
Under our assumption $a_{\mu\; a}=0$, one can justify to set
$v_{(i\vec{m})} = 0$ also from (\ref{eq:F-equations3}) as well as
(\ref{eq:DF-impose}) by an analogous discussion.}.
Such zero modes
associated with Abelian sub-algebra do not couple
to the interaction.
Moreover, as we have seen, 
$v_{(i\vec{m})}$ can be absorbed into the shift of the $X^I_{(i\vec{m})}$
and then is irrelevant for $X^I$ and $\Psi$ part Lagrangian.
Therefore, as a solution to the equations,
we can set $v_{(i\vec{m})} = 0$.

After setting $a_{\mu\; a} = v_{(i\vec{m})} = 0$,
by using
(\ref{eqs:relation-A1})-(\ref{eqs:relation-A5})
we find that the non-zero components of $\left( D^\mu \tilde{F}_{\mu\nu}
\right)^A{}_{B}$ are
\begin{align}
\label{eq:DF-1}
    \left( D^\mu \tilde{F}_{\mu\nu} \right)^0{}_{(i\vec{m})}=&
i m^a v_a 
\left(\hat{D}^\mu {F}_{\mu\nu} \right)_{(i\vec{m})}
\,,
\\
\label{eq:DF-2}
    \left( D^\mu \tilde{F}_{\mu\nu} \right)^a{}_{(i\vec{m})}=&
- i  m^a
\left(\hat{D}^\mu {F}_{\mu\nu} \right)_{(i\vec{m})}
\,,
\\
\label{eq:DF-3}
    \left( D^\mu \tilde{F}_{\mu\nu} \right)^{(i\vec{m})}{}_{(j\vec{n})}=&
- f^{ki}{}_j
\left(\hat{D}^\mu {F}_{\mu\nu} \right)_{(k,\vec{n}-\vec{m})}
\end{align}
where we have defined
\begin{align}
F_{\mu\nu\; (i\vec{m})} =&
\left(
\partial_\mu A_\nu
-\partial_\nu A_\mu
+i \commt{A_\mu}{A_\nu}
\right)_{(i\vec{m})} \,.
\end{align}
Here the definition of the covariant derivative $\hat{D}_{\mu}$ is
given in (\ref{eq:def_Hat_D}):
\begin{equation}
(\hat{D}^{\mu}F_{\mu\nu})_{(i\vec{m})} 
=
\partial^{\mu} F_{\mu\nu\;(i\vec{m})}
+i [A^{\mu}, F_{\mu\nu}]_{(i\vec{m})} 
.
\end{equation}
Then 
by comparing these
with 
(\ref{eq:DF-current1})-(\ref{eq:DF-current5}),
we have 
a single equation of motion for the non-vanishing gauge field:
\begin{align}
\left( \hat{D}^\mu F_{\mu\nu} \right)_{(i\vec{m})} =
\tilde{J}_{\nu\; (i\vec{m})} \,,
\end{align}
where $\tilde{J}_{\nu\; (i\vec{m})} =  C^\mu_0 J_{\mu\nu\; (i\vec{m})}$.

Now one can check that the current term is derived from
$\mathcal{L}_{X+\Psi}$ in (\ref{eq:gen_C_XPsi_Lag}):
\begin{align}
  \frac{\delta \mathcal{L}_{X+\Psi}}{\delta
    A^\nu_{(i,-\vec{m})}}
=&
-g_{ab}\partial^{a}(\hat{D}_{\nu}Y^b -\partial^b A_{\nu})_{(i\vec{m})}
-iP^{I}_{J}\commt{{X}_I}{\hat{D}_\nu X^J}_{(i\vec{m})}
\nn \\ 
&-ig_{ab}\commt{{Y}^a}{(\hat{D}_\nu Y^b - \partial^b A_{\nu} )}_{(i\vec{m})}
+\frac{1}{2} \commt{\Bar\Psi}{\Gamma_\nu \Psi}_{(i\vec{m})}
\nn\\=&
-\frac{1}{C^2} \tilde{J}_{\nu\; (i\vec{m})} \,.
\end{align}
Therefore, the equation of motion can be derived from  the standard 
Lagrangian, $-\frac{1}{4C^2} F_{\mu\nu\; (i,-\vec{m})}
F^{\mu\nu}_{(i\vec{m})}$.

\paragraph{Supersymmetry}

We 
 look at the condition for supersymmetry to be preserved under
a given set of the moduli.
We have now,
\begin{align}
    X_0^I = \lambda_0^I \,, 
\quad
 X_a^I = \lambda_a^I \,,
\quad
C_0^{\tilde\mu}, \,
H_{\mu\nu\rho\, 0} , \, H_{\mu\nu\rho\, a} = \text{const.} \,,
\quad
C_a^{\tilde\mu} = v_a C_0^{\tilde\mu} \,,
\end{align}
and 
the others are zero.
Then the supersymmetry transformation of each component of $\Psi$ is
\begin{align}
    \delta \Psi_{\ul{\alpha}} = 0 \,,
\qquad
\delta \Psi_{(i\vec{m})} = 0
\,,
\qquad
\delta \Psi_{\alpha} =
\frac{1}{3!}\frac{1}{2}\Gamma^{\mu\nu\rho} H_{\mu\nu\rho\, \alpha}
\,,
\end{align}
where $\alpha= (0,a)$.
Note that $\delta \Psi_{(i\vec{m})}=0$ since all
gauge fields $\AAB{\mu}{B}{A}$ are set to be zero
as the background.
These relations mean that our choice of the moduli does not break
supersymmetry
if we take $H_{\mu\nu\rho\, 0} = H_{\mu\nu\rho\, a}=0$.
Since these components of three-form do not appear in the effective
action, namely they are decoupled, then we will set them to
vanish.

\paragraph{Summary}

In summary, we write down the effective Lagrangian in this case:
\begin{align}
{\cal L}=&
-\frac{1}{2} P_{IJ}
\left[
   \left( \hat{D}^{\hat\mu} X^I \right)_{(i,-\vec{m})}
   \left( \hat{D}_{\hat\mu} X^J \right)_{(i\vec{m})}
+C^2 g_{ab} 
   \left( \hat{D}^a X^I \right)_{(i,-\vec{m})}
   \left( \hat{D}^b X^J \right)_{(i\vec{m})} \right]
\nn\\ &
+\frac{1}{4}C^2 P_{IK} P_{JL} 
  \commt{X^I}{X^J}_{(i,-\vec{m})}
  \commt{X^K}{X^L}_{(i\vec{m})}
\nn\\ &
+\frac{i}{2} \Bar\Psi_{(i,-\vec{m})}
 \left( \Gamma^{\hat\mu} ( \hat{D}_{\hat\mu} \Psi )
-C_0^{\tilde\mu} \Gamma_{\tilde\mu} \Gamma^I \tau_a^I ( \hat{D}^a \Psi )
 \right)_{(i\vec{m})} 
\nn\\ &
+\frac{1}{2} \Bar\Psi_{(i,-\vec{m})}
 C_0^{\tilde\mu} \Gamma_{\tilde\mu} 
 \Gamma^I  P^I_J \commt{X^J}{\Psi}_{(i\vec{m})}    
\nn\\ &
-\frac{1}{4 C^2}
\left[
F_{{\hat\mu}{\hat\nu}\; (i,-\vec{m})} F^{{\hat\mu}{\hat\nu}}_{(i\vec{m})}
+ C^4 g_{ac} g_{bd} 
  F^{ab}_{(i,-\vec{m})} F^{cd}_{(i\vec{m})}
+2 C^2 g_{ab}\eta_{{\hat\mu}{\hat\nu}}
F^{a{\hat\mu}}_{(i,-\vec{m})} F^{b{\hat\nu}}_{(i\vec{m})}
\right]
\,,
\end{align}
where we rewrite $f^{ab}$ as
$F^{ab}$, and define a dimensionless gauge field
$A^a_{(i\vec{m})} = Y^a_{(i\vec{m})}$ and 
covariant derivatives
\begin{align}
    \left( \hat{D}_{\hat\mu} \phi \right)_{(i\vec{m})}
=& \partial_{\hat\mu} \phi_{(i\vec{m})} + i \commt{A_{\hat\mu}}{\phi}_{(i\vec{m})}
\,,
\\
    \left( \hat{D}_a \phi \right)_{(i\vec{m})}
=& (\partial_a \phi)_{(i\vec{m})} + i \commt{A_a}{\phi}_{(i\vec{m})}
\,.
&(\partial_a = i m_a)
\end{align}
$a_{\mu \; a}$ and $v_{(i\vec{m})}$ have disappeared.
At this stage, we have taken into account
the effect of the dimensional reduction
from (\ref{eq:dim_rec_byC}), and
$\hat\mu$ and $\hat\nu$ denote other directions than $\tilde\mu$,
namely $\hat\mu=0,\cdots ,4$ if $\tilde\mu=5$.
Note that the field strength is also restricted due to the condition
(\ref{feq4-2}).

This Lagrangian can be seen as supersymmetric Yang-Mills theory
defined on $\bR^{1,4} \times T^d$.
Now one of the $\mu$ direction, $\mu=\tilde{\mu}$, is reduced by 
constraint.
Instead, we have added world-volume directions equipped with the metric
$g_{ab}$.
The $C$-field is completely a constant.
There are two other parameters,
$v_a$ and $v_{(i\vec{m})}$, which are proportional constants of
$C_a^{\tilde\mu}$ and $C_{(i\vec{m})}^{\tilde\mu}$ components to
$C_0^{\tilde\mu}$
respectively.
$v_{(i\vec{m})}$ is set to be zero as a result of the equations of
motion,
and $v_a$ is constant and is
combined with the moduli parameters $\lambda_0^I$ and
$\lambda_a^I$, which are from $X^I$, to form a ``vielbein'' 
$\tau_a^I = \lambda_a^I - v_a \lambda_0^I$.
The internal metric is given by $g_{ab} = \tau_a^I \tau_b^I$.
An appropriate combination of $(C_0^{\tilde\mu})^2$ and $g_{ab}$
indeed gives the
volume of the torus $T^d$.
$C_0^{\tilde\mu}$ appears solely as a coefficient of interaction terms,
and we can 
view the role of 
$C_0^{\tilde\mu}$ as a coupling constant.
We well concretely see the relation in the case of $d=1$ later.


\subsection{$C_0=0$ case}
\label{sec:C0_zero}

In this section, we consider the case with $C_0^\mu =
C_{(i\vec{m})}^\mu=0$, which solves the condition $C_C^\mu C_D^\nu
{f^{CDB}}_{A}=0$.
Then this case is the case 3 of section \ref{sec:constraints},
$C^\mu = C_a^\mu u^a$.
As we will see, from this setup we have a different kind of action on
a torus.

\paragraph{$C$ fields and gauge fields}

As mentioned above, we consider the case,
\begin{align}
    C_0^\mu =0 \,,
\qquad
C_{(i\vec{m})}^\mu =0 \,,
\qquad
C_a^\mu \neq 0 \,.
\end{align}
This condition trivially solves one of the constraint equations (\ref{feq5}),
$C_A^\mu C_B^\nu f^{ABC}{}_D=0$.
We then start with the other of (\ref{feq5}), $(D_\mu C^\nu)_A=0$.
For $A=\ul{0}, \ul{a}$, this becomes the ghost decoupling condition as
before.
Likewise the previous case, we again impose the condition that
$\partial_{\nu}C_{\ul{0}}^\mu = \partial_{\nu}C_{\ul{a}}^\mu =0$ for decoupling of the ghost
modes.
For $A=a$, this condition means $\partial_\mu C^\nu_a =0$, namely, the
non-zero components of $C_a^\mu$ have to be constant.
For $A=0, (i\vec{m})$, this becomes
\begin{align}
    \AAB{\mu}{a}{0} C_a^\nu = \AAB{\mu}{a}{(i\vec{m})} C_a^\nu =0 \,.
\end{align}
Thus for $a$ with non-zero $C_a^\mu$,
$\AAB{\mu}{a}{0}=\AAB{\mu}{a}{(i\vec{m})}=0$.
Without tilde, this condition implies
\begin{align}
    A_{\mu\; (i\vec{m})(i,-\vec{m})} f^{(i\vec{m})(i,-\vec{m})0}{}_a
      =0
\,,
\qquad
A_{\mu\; 0(i\vec{m})}f^{0(i\vec{m})a}{}_{(i\vec{m})} =0 \,. 
\end{align}
Since the structure constant here is proportional to $m^a$, we have
the condition, $A_{\mu\; (i\vec{m})(i,-\vec{m})}$ for all $\vec{m}$,
and $A_{\mu\; 0(i\vec{m}\neq \vec{0})}=0$, but
$A_{\mu\; 0(i\vec{0})}$ is unconstrained.
Therefore the non-vanishing gauge fields are
\begin{align}
\label{tildeA=A-Ca}
    \AAB{\mu}{0}{(i\vec{m})} =& -i m^a A_{\mu\; a(i\vec{m})} +
    f^{jk}{}_i A_{\mu\; (j\vec{n})(k,\vec{m}-\vec{n})} \,,
\nn\\
\AAB{\mu}{(i\vec{m})}{(i\vec{m})} =& -i m^a A_{\mu\; 0a} \,,
\nn\\
\AAB{\mu}{(i\vec{m})}{(j\vec{n})} =& f^{ki}{}_j A_{\mu\; 0 (k\vec{0})}
\delta^{\vec{m}-\vec{n}} \,.
\end{align}
With these gauge fields, the covariant derivatives are
\begin{align}
    \left(D_\mu \phi \right)_0 =& \partial_\mu \phi_0 \,,
\quad
    \left(D_\mu \phi \right)_a = \partial_\mu \phi_a \,,
\quad
    \left(D_\mu \phi \right)_{\ul{a}} = \partial_\mu \phi_{\ul{a}} \,,
\nn\\
    \left(D_\mu \phi \right)_{\ul{0}} =&
 \partial_\mu \phi_{\ul{0}} + \AAB{\mu}{0}{(i,-\vec{m})}
 \phi_{(i\vec{m})} \,,
\nn\\
    \left(D_\mu \phi \right)_{(i\vec{m})} =&
    \left(\tilde{D}_\mu \phi \right)_{(i\vec{m})}
 - \AAB{\mu}{0}{(i\vec{m})}
 \phi_{0} \,.
\end{align}

\paragraph{$X$ and $\Psi$ part}

For the bosonic field $X^I_A$ and the fermionic field $\Psi_A$,
we impose the same ghost decoupling condition as before.
Namely, we gauge away $u^{\ul{0}}$ and $u^{\ul{a}}$ components and
take the conjugate components of $X^I$ and $\Psi$ to be moduli:
\begin{align}
    X^I_0 =& \lambda_0^I \,,
\qquad
X^I_a = \lambda_a^I \,,
\qquad
\Psi_0 = \Psi_a = 0 \,,
\end{align}
where $\lambda_0^I$ and $\lambda_a^I$ are constant.
As we will see, there will not appear $\lambda_a^I$ in the effective
equations of motion below, and then it is sufficient to consider
$\lambda_0^I$ only.
We then use $\lambda^I \equiv \lambda_0^I$ in this subsection.
We can define the projector as before, which in this case takes
the form:
\begin{align}
    P^I_J =& \delta^I_J - \frac{\lambda^I \lambda_J}{\lambda^2} \,,
\qquad
P^I_J \lambda^J = \lambda_I P^I_J = 0 \,,
\end{align}
and by using this we decompose 
$X^I_{(i\vec{m})}$ as
$X^I_{(i\vec{m})} = P^I_J X^J_{(i\vec{m})} + \lambda^I Y_{(i\vec{m})}$
where $Y_{(i\vec{m})} = \lambda_J X^J_{(i\vec{m})} / \lambda^2$.
Note that since we are considering constant $\lambda^I$, we can
choose one direction in which $\lambda^I$ is non-vanishing, for example,
$\lambda^I= \lambda^{10} \delta^{I 10}$.
For this case the projector selects the directions $\hat{I}=6,7,8,9$.
We will revisit this point in the summary part of this subsection.

After decoupling  the ghost part, $X$ and $\Psi$ equations of motion are
given through the generic Lagrangian
(\ref{eq:Lag_X_gen}) and (\ref{eq:Lag_Psi_gen}) again.
We substitute our ansatz into them, and then obtain 
\begin{align}
\label{eq:Lag_X_Psi_Ca}
    \mathcal{L}_{X+\Psi}=&
-\frac{1}{2} P_{IJ} \left(
\left( \tilde{D}_\mu X^I \right)_{(i,-\vec{m})}
\left( \tilde{D}^\mu X^J \right)_{(i\vec{m})}
+\lambda^2 \tilde{g}_{ab} 
\left( \partial^a X^I \right)_{(i,-\vec{m})}
\left( \partial^b X^J \right)_{(i\vec{m})}
\right)
\nn\\ & 
-\frac{\lambda^2}{2} 
\left(
\left( \tilde{D}_\mu Y \right)_{(i,-\vec{m})} - \AAB{\mu}{0}{(i,-\vec{m})}
 \right) 
\left(
\left( \tilde{D}^\mu Y \right)_{(i\vec{m})} - \tilde{A}^{\mu\; 0}{}_{(i\vec{m})}
 \right)
\nn\\ &
+\frac{i}{2} \Bar\Psi_{(i,-\vec{m})} \Gamma^\mu
\left(
\tilde{D}_\mu \Psi
+  \lambda^I \Gamma^I C_{\mu\; a} \partial^a \Psi
\right)_{(i\vec{m})} \,,
\end{align}
where $\tilde{g}_{ab} \equiv C^\mu_a C_{\mu\; b}$ and again
$\partial^a = i m^a$.
Therefore in this case we do not have potential terms for $X^I$ and $\Psi$.

\paragraph{Gauge field part}

Also in this case, the gauge field equation of motion is given through
(\ref{feq2}) and (\ref{feq4}) by (\ref{eq:DF=J_rel}):
\begin{align}
      \left( D^\mu \tilde{F}_{\mu\nu} \right)^B{}_A
=&
C_C^\lambda J_{\nu\lambda\; D} f^{CDB}{}_A \,,
\end{align}
where the ``current'' $J_{\nu\lambda\; A}$ is again given by
(\ref{eq:Def_J}):
\begin{align}
      J_{\nu\lambda\; A} =&
2i \thcommt{C_{[\nu}}{X^I}{D_{\lambda]} X^I}_A
- \thcommt{C_{[\nu}}{\Bar\Psi}{\Gamma_{\lambda]} \Psi}_A \,.
\end{align}
As before, only $J_{\nu\lambda\; (i\vec{m})}$ components contributes
to the equation of motion.
The ghost part of $H_{\mu\nu\rho\; \ul{0}}$ and $H_{\mu\nu\rho\;
  \ul{a}}$ are taken care of in the same way as before.
Since now only $C_a^\mu$ is non-zero, the only non-trivial equation of
motion is
\begin{align}
    \left( D^\mu \tilde{F}_{\mu\nu} \right)^0{}_{(i\vec{m})} =&
-i m^b C_b^\lambda J_{\nu\lambda\; (i\vec{m})}
\nn\\=&
\lambda^2 \tilde{g}_{ab} m^a m^b \left(
(\tilde{D}_\nu Y)_{(i\vec{m})}
-\AAB{\nu}{0}{(i\vec{m})}
\right) \,,
\end{align}
and the other components are 
$\left( D^\mu \tilde{F}_{\mu\nu} \right)^A{}_B =0$.

By defining $A_{\mu \; 0(k \vec{0})} = A^{(0)}_{\mu\; k}$ and
$A_{\mu \; 0a} = -a_{\mu\; a}$, the field strength can be written down
as
\begin{align}
\FAB{\mu}{\nu}{(i\vec{m})}{(j\vec{n})} =
-f^{ki}{}_j F^{(0)}_{\mu\nu\; k} \delta^{\vec{m}}_{\vec{n}} \,,
\qquad
\FAB{\mu}{\nu}{(i\vec{m})}{(i\vec{m})} =
-i m^a f_{\mu\nu\; a} \,,
\end{align}
where $F^{(0)}_{\mu\nu\; k} = 2 \partial_{[\mu} A^{(0)}_{\nu]\; k} 
-2f^{ij}{}_k A^{(0)}_{[\mu \; i} A^{(0)}_{\nu] \; j}$,
$f_{\mu\nu \; a} = 2\partial_{[\mu} a_{\nu] \; a}$.
$\FAB{\mu}{\nu}{0}{(i\vec{m})}$ is also non-zero and depends on
all the gauge fields $A^{(0)}_{\mu\; k}$, $a_{\mu\; a}$ and
$\AAB{\mu}{0}{(i\vec{m})}$.
The rest of the gauge field strengths, including $\FAB{\mu}{\nu}{0}{\ul{a}}$
and $\FAB{\mu}{\nu}{a}{(i\vec{m})}$, all vanish.
We start with the equations of motion with $(i\vec{m})(j\vec{n})$ index
and $(i\vec{m})(i\vec{m})$ index.
As mentioned above, there is no source term for these components and
the equations of motion are
\begin{align}
\label{eq:EOM_A0-a_Ca}
    -f^{ki}{}_j \left( \partial^\mu F^{(0)}_{\mu\nu\; k} 
- 2f^{st}{}_k A^{(0)\; \mu}_{\ph{\mu} \; [s} F^{(0)}_{t]\;\mu\nu}
\right) =0 \,,
\qquad
-i m^a \partial^\mu f_{\mu\nu \; a} =0 \,. 
\end{align}
On the other hand, if we differentiate the effective Lagrangian
(\ref{eq:Lag_X_Psi_Ca}) for $X^I$ and $\Psi$ after the gauge fields
$A^{(0)}_{\mu \; k}$ and $a_{\mu\; a}$, it is easy to see that we have
non-zero currents consisting of $X^I$ and $\Psi$.
So these gauge fields have to be regarded as backgrounds, so that
we do not consider the variations of these fields in the action.
Since the background gauge fields satisfy the equations of motion without
sources (\ref{eq:EOM_A0-a_Ca}), we will assume the simplest solution
$A^{(0)}_{\mu \; k} = a_{\mu\; a}=0$.
Note that this choice makes the covariant derivative $\tilde{D}_\mu$
appearing in the action (\ref{eq:Lag_X_Psi_Ca}) the ordinary one $\partial_\mu$.
Finally, the nontrivial equation of motion for the gauge field 
becomes
\begin{align}
\label{eq:EOMgauge-Ca}
    \partial^\mu \FAB{\mu}{\nu}{0}{(i\vec{m})} =&
\lambda^2 \tilde{g}_{ab} m^a m^b 
\left( \partial_\nu Y_{(i\vec{m})} - \AAB{\nu}{0}{(i\vec{m})} \right)
\nn\\=&
\tilde{g}_{ab} m^a m^b \frac{\delta \mathcal{L}_{X+\Psi}}{\delta
  \tilde{A}^{\nu \; 0}{}_{(i,-\vec{m})}} \,,
\end{align}
and $\FAB{\mu}{\nu}{0}{(i\vec{m})} =
 -\partial_\mu \AAB{\nu}{0}{(i\vec{m})}+\partial_\nu \AAB{\mu}{0}{(i\vec{m})}$
is 
an Abelian field strength.
Therefore we just have copies of the Abelian gauge field.

Note that the gauge field $\AAB{\mu}{0}{(i\vec{m})}$ appears either
in the combination $\partial_\nu Y_{(i\vec{m})} -
\AAB{\nu}{0}{(i\vec{m})}$
or in the field strength $\FAB{\mu}{\nu}{0}{(i\vec{m})}$.
Let us consider the residual gauge transformation, with the gauge
parameter $\tilde\Lambda^0{}_{(i\vec{m})}$, of the fields.
It is easy to see that $P^I_J X^J_{(i\vec{m})}$ and
$\Psi_{(i\vec{m})}$ do not transform under this gauge transformation,
while $Y_{(i\vec{m})}$ transforms as 
$Y_{(i\vec{m})} \rightarrow Y_{(i\vec{m})} + \tilde\Lambda^0{}_{(i\vec{m})}$.
We have set the background field condition,
$A^{(0)}_{\mu \; k} = a_{\mu\; a}=0$.
Next we write the term involving $Y_{(i\vec{m})}$
in the effective Lagrangian~(\ref{eq:Lag_X_Psi_Ca})
in the Fourier transformed form:
\begin{equation}
-\frac{\lambda^2}{2} 
\int_0^{2\pi} \frac{d^d y}{(2\pi)^d} \,
\left(
\partial_\mu Y_i(y) - A_{\mu\; i}(y)
 \right) 
\left(
\partial^\mu Y_i(y) - A^{\mu}_{\; i}(y)
 \right) \,,
\end{equation}
where $Y_i(y) = \sum_{\vec{m}} Y_{(i\vec{m})} e^{i \vec{m} \cdot
  \vec{y}}$
and $A_{\mu\; i}(y) = \sum_{\vec{m}} \AAB{\mu}{0}{(i\vec{m})} e^{i \vec{m} \cdot
  \vec{y}}$,
and the suffix $i$ is just the label for
Abelian fields of which we now have a number of copies.
Then this term turns out to be  written in the form of
the $U(1)$ complex Higgs kinetic term as
\begin{equation}
-\frac{1}{2}
\int_0^{2\pi} \frac{d^d y}{(2\pi)^d} \,
\left( {\cal D}_{\mu}\varphi_i \right)^\dagger
\left( {\cal D}^{\mu}\varphi_i \right) \,,
\label{eq:Higgs}
\end{equation}
where $\varphi_i(y) = \sqrt{|\lambda|^2} e^{iY_i(y)}$, and
the covariant derivative is 
${\cal D}_{\mu}\varphi_i
= \partial_{\mu} \varphi_i - i A_{\mu\; i} \varphi_i$,
where the index $i$ is not summed over.
Note that $Y_i(y)$ is dimensionless
since $\lambda^I Y_{(i\vec{m})}$ has the same mass dimension as
$\langle X^I_{0} \rangle = \lambda^I$.
The fluctuation with respect to the magnitude $|\varphi_i|$
comes from the fluctuation of $X^I_0$.
This fluctuation is suppressed as a result of the ghost decoupling 
discussed in section~\ref{sec:ghost_decoupling}.
So there is no fluctuation along the absolute value.

The part (\ref{eq:Higgs}) has the $U(1)$ gauge symmetries as 
\begin{equation}
\varphi_i \to e^{i\Lambda_i(y)}\varphi_i 
\,, \label{U1} 
\end{equation}
with the Fourier transformed gauge parameters $\Lambda_i(y)$.
Now the action
(\ref{eq:Lag_X_Psi_Ca})
is the expansion of the (\ref{eq:Higgs}) around the 
following vacuum expectation values  (VEVs)\footnote{%
Though we only deal with the classical equations of motion in this
paper, we abuse the term VEV to refer to constant solutions to the equations
of motion, around which we derive new equations of motion for the
dynamical fields.
This would not cause any confusion.},
\begin{equation}
\langle \varphi_i \rangle = \sqrt{|\lambda |^2}, \qquad
\langle Y_{i} \rangle = 0. 
\end{equation}
These VEVs break the 
$U(1)$ gauge transformations
(\ref{U1}).
Then we can regard $Y_i(y)$ as the 
Goldstone modes along the broken $U(1)$ direction.
Since $Y_i(y)$ are the Goldstone modes,
like the usual Higgs mechanism,
we can absorb these modes by redefining
the gauge fields $A_i(y)$.
In terms of the original Fourier basis,
we define
\begin{equation}
w_{\mu\; (i\vec{m})} = -\AAB{\mu}{0}{(i\vec{m})} + \partial_\mu
Y_{(i\vec{m})}  \,.
\end{equation}
Then the field strength is now written in terms of $w_{\mu\;
  (i\vec{m})}$:
\begin{align}
\FAB{\mu}{\nu}{0}{(i\vec{m})}
=& -\left( \partial_\mu \AAB{\nu}{0}{(i\vec{m})} 
- \partial_\nu  \AAB{\mu}{0}{(i\vec{m})} \right) \nn \\
=& \left( \partial_\mu w_{\nu} - \partial_\nu w_\mu
   \right)_{(i\vec{m})} \nn \\
=& W_{\mu\nu\; (i\vec{m})} \,.
\end{align}
The $w_{\mu\; (i\vec{m})}$ are the massive gauge bosons
which already absorb the Goldstone modes $Y_{(i\vec{m})}$,
and the equations of motion for $Y_{(i\vec{m})}$ and
$\AAB{\mu}{0}{(i\vec{m})}$ part can be obtained from
the following W-part Lagrangian:
\begin{equation}
    {\cal L}_W = 
-\frac{\lambda^2}{2} \bar{g}_{ab} m^a m^b w^\mu_{(i,-\vec{m})} w_{\mu\; (i\vec{m})}
-\frac{1}{4} W^{\mu\nu}_{(i,-\vec{m})} W_{\mu\nu \; (i\vec{m})} \,.
\end{equation}
The first term can be seen as the mass term for W-bosons
produced by the $U(1)$ breaking.
The number of the independent polarization
for each $w_{\mu\; (i\vec{m})}$ is
$6-d-2+1=5-d$
\footnote{%
After the projection, the number of the transverse bosons is 4 and then
the total number of bosonic degrees of freedom is now $9-d$, while the
fermionic one is $8$. 
Therefore, the effective Lagrangian may not be supersymmetric except
for $d=1$ case.
In the case of $d=1$, the supersymmetry is similar to D4-brane's because of the dimensional reduction.}.
Here $-d$ is due to the constraint (\ref{feq6}),
$-2$ is for elimination of the temporal and the longitudinal modes,
and $+1$ is from 
absorption of $Y$ boson.

Next let us consider the geometrical meaning of this Higgs mechanism
in terms of the target space description. 
This Higgs mechanism eliminates the one of the transverse directions
from the action,
and then can be considered 
as the dimensional reduction of M-theory
to type IIA string theory.
We identify 
$\sqrt{|\lambda |^2}$ as the radius of the 
circle, and the phase $Y_{(i\vec{m})}$ of 
$\varphi$ as the coordinate along the M-circle.
The VEV $\langle \varphi \rangle = \sqrt{|\lambda |^2}$
as well as 
$\langle Y_{(i\vec{m})} \rangle = 0$ represents the 
position of the 5-brane in the compactified direction,
and the 5-brane breaks the shift symmetry 
along the compactified direction.
Now because of the projection, $Y_{(i\vec{m})}$ enjoys not only the
global shift symmetry but also the gauged one,
namely $U(1)$ symmetry. 
As we will see in section \ref{sec:vanishing_C}, the gauge field
$\AAB{\mu}{0}{(i\vec{m})}$ can be viewed as the background graviphoton
field arising from the compactification.
Therefore, it is natural that these gauge fields, corresponding to the
local reparametrization on the circle, absorb the Goldstone modes
and become massive.
As a result, we have an effective Lagrangian of
five-brane in string theory.

Since there is a constraint:
\begin{eqnarray}
m^aC_a^{\mu}(D_{\mu}\phi)_{(i\vec{m})}=0 \,, \label{constraint}
\end{eqnarray}
one might wonder if the theory is capable of realizing
the target space whose dimensions less than ten dimensions.
But the theory 
remains ten-dimensional
even under the constraint
(\ref{constraint}), as we now observe.
At the first sight, (\ref{constraint})
prohibits the covariant derivative along the direction of $C^{\mu}$.
The number of the target space dimensions
$\bR^{1,9}$
and the world-volume dimensions $\bR^{1,5}$  are reduced as
$\bR^{1,9} \to \bR^{1,9-d}$ and $\bR^{1,5} \to \bR^{1,5-d}$
respectively.
But the reduced directions are recovered by the KK-momentum
$C^{\mu}_{a}\partial^a$, and thus
the actual target space and the world-volume are
$\bR^{1,9-d} \times T^d$ and 
$\bR^{1,5-d} \times T^d$ respectively.
So the theory remains to be a 5-brane effective theory 
of a (1+9)-dimensional superstring theory, irrespective of how many
$C_a^\mu$ we have turned on.

\paragraph{Summary}

We have the effective Lagrangian:
\begin{align}
\label{eq:Lag_Ca_1}
    {\cal L} =&
-\frac{1}{2} P_{\hat{I}\hat{J}} \left(
\left( \partial_\mu X^{\hat{I}} \right)_{(i,-\vec{m})}
\left( \partial^\mu X^{\hat{J}} \right)_{(i\vec{m})}
+\lambda^2 \tilde{g}_{ab} 
\left( \partial^a X^{\hat{I}} \right)_{(i,-\vec{m})}
\left( \partial^b X^{\hat{J}} \right)_{(i\vec{m})}
\right)
\nn\\ &
+\frac{i}{2} \Bar\Psi_{(i,-\vec{m})} \Gamma^\mu
\left(
\partial_\mu \Psi
+  \lambda^I \Gamma^I C_{\mu\; a} \partial^a \Psi
\right)_{(i\vec{m})}
\nn\\ &
-\frac{\lambda^2}{2} \tilde{g}_{ab} m^a m^b 
  w^\mu_{(i,-\vec{m})} w_{\mu\; (i\vec{m})}
-\frac{1}{4} W^{\mu\nu}_{(i,-\vec{m})} W_{\mu\nu \; (i\vec{m})} \,,
\end{align}
where $\tilde{g}_{ab} = C^\mu_a C_{\mu\; b}$.
$\lambda_a^I=X_a^I$ do not show up here and can then be set to zero.
For simplicity,
$\lambda^I=X_0^I$ is chosen as $\lambda^I=\lambda^{10} \delta^{I 10}$
and then $P^I_J$ is the projection onto $\hat{I}=6,7,8,9$ plane.
The world volume is ${\bf R}^{1,5-d}\times T^d$.
Indices $\mu, \nu$
label ${\bf R}^{1,5-d}$ directions, and $a,b$ label $T^d$
directions.
This Lagrangian might be able to couple to background gauge fields
$A^{(0)}_{\mu\; i}$ and $a_{\mu\; a}$ by replacing the derivative with
covariant derivatives, but now the background fields are turned off.
$w_\mu$ field is defined by
\begin{align}
    w_{\mu\; (i\vec{m})} =& -\AAB{\mu}{0}{(i\vec{m})} + \partial_\mu
    Y_{(i\vec{m})}
\,,
\end{align}
and thus this Lagrangian is a gauge fixed Lagrangian with massive
Abelian vector bosons.
All the fields have the Kaluza-Klein mass term whose mass is
determined by $\lambda^2 \tilde{g}_{ab}$
\footnote{ 
The Higgs mechanism in this section is a kind of St\"ueckelberg mechanism 
\cite{Stueckelberg:1900zz}
(for a review, see \cite{Ruegg:2003ps}).
It has been known that non-Abelian extension of St\"ueckelberg
mechanism has the problems of renormalizability and unitarity in
4-dimension.
In this paper, the effective Lagrangian is reduced to an Abelian
system, regardless of the 3-algebraic structure in the formulation,
and then there will not be such problems.}.

The theory is regarded as the effective theory of Abelian 5-branes
in a superstring theory, which is
similar to the D5-brane or NS5-brane effective theory
in type IIB superstring theory.
Moreover we have seen that the compactification to string theory
occurs along the 
direction transverse to the 5-brane world-volume,
and such compactification usually gives NS5-branes in type IIA string
theory.
However this 5-brane action should be 
recognized as a
 type IIB NS5-brane action derived from the 
type IIA NS5-brane by T-duality,
since the KK-momentum shows up along the 
5-brane world-volume directions.
This theory is interesting since it captures the 
history of the 5-branes generated through the 
M-theory compactification of  M5-branes 
and T-duality of type IIA NS5-branes.


\subsection{Vanishing $C$ field case}
\label{sec:vanishing_C}

In this subsection, we consider the simplest case, 
i.e.,
all $C_A^\mu$ vanish.
This is the case 4 in section \ref{sec:constraints}.
In this case, $\FAB{\mu}{\nu}{B}{A}=0$ because of 
(\ref{feq4}); that is, the auxiliary gauge
field $\AAB{\mu}{B}{A}$ is a pure gauge, but still couples to the
other fields through the covariant derivatives.
The equations of motion are reduced to
\begin{align}
\label{eq:EOM_C=0}
    D^2 X^I_A =0 
\,,
\quad
D_{[\mu}H_{\nu\lambda\rho]\; A} =0
\,,
\quad
\Gamma^\mu (D_\mu \Psi)_A =0 \,.
\end{align}
Apart from the covariant derivative, these are just the equations of
motion for Abelian $(2,0)$ tensor multiplets in six dimensions.
Therefore, we may assume the Lagrangian \textit{\'a la}
Pasti-Sorokin-Tonin (PST~\cite{Pasti:1997gx}):
\begin{align}
\label{eq:Lag_PST_1}
    {\cal L} =&
-\frac{1}{2} \left\langle
\left(D^\mu X^I \right) ,
\left(D_\mu X^I \right) \right\rangle
+\frac{i}{2} \left\langle
\Bar\Psi ,
\Gamma^\mu \left( D_\mu \Psi \right)
\right\rangle
+\frac{1}{4}
\left\langle
 H^*_{\mu\nu} ,
\left( H^{* \; \mu\nu} - H^{\mu\nu} \right)
\right\rangle \,,
\end{align}
where
\begin{align}
    H_{\mu\nu\; A} =&
 \frac{\partial^\rho a}{\sqrt{\partial_\mu a \partial^\mu a}} H_{\mu\nu\rho \; A} \,,
\qquad
    H^*_{\mu\nu\; A} =
 \frac{\partial^\rho a}{\sqrt{\partial_\mu a \partial^\mu a}} 
\frac{\epsilon_{\mu\nu\rho\lambda\sigma\tau}}{3!} H^{\lambda\sigma\tau}{}_A \,,
\end{align}
and now the three-form 
can be written in terms of a two-form potential $b_{\mu\nu\; A}$ as 
$H_{\mu\nu\rho\; A}= 3 D_{[\mu} b_{\nu\rho]\; A}$,
thanks to the usual Bianchi identity and $\FAB{\mu}{\nu}{B}{A}=0$.
Note that the three-form $H_{\mu\nu\rho\; A}$ is assumed to be not
self-dual.
$a$ is an auxiliary scalar field of PST action
and is singlet under the three algebraic transformation,
and with it the effective Lagrangian (\ref{eq:Lag_PST_1})
enjoys the
following local symmetries:
\begin{align}
(\text{I}):& \quad
  \delta_\text{I} a =0 \,,
\qquad
  \delta_\text{I} b_{\mu\nu\; A} = 2 D_{[\mu} \xi_{\nu]\; A} \,,
\nn\\  
(\text{II}):& \quad
  \delta_\text{II} a =0 \,,
\qquad
  \delta_\text{II} b_{\mu\nu\; A} = 2 \partial_{[\nu} a \,  \eta_{\mu ] \; A}\,,
\nn\\
(\text{III}):& \quad
  \delta_\text{III} a =\zeta \,,
\qquad
  \delta_\text{III} b_{\mu\nu\; A}
= \frac{\zeta}{\sqrt{(\partial a)^2}} \left(
H^*_{\mu\nu\; A} - H_{\mu\nu\; A}
\right) \,.
\end{align}
The first transformation $\delta_\text{I}$ agrees with
 the usual gauge symmetry of the $b_{\mu\nu\; A}$ field
because of $\commt{D_\mu}{D_\nu}=0$.
The second and the third symmetries are characteristic for the PST
formalism
and are important for the three-form to be on-shell self-dual.
Therefore, the effective action can reproduce the equations of motion
(\ref{eq:EOM_C=0}), including the linear self-duality condition.

Likewise the previous cases, we gauge the shift symmetry and then
gauge away the ghost modes, $X^I_{\ul{0}}$ etc.
In this case, we have the Lagrangian description for the three-form
field strength $H_{\mu\nu\rho\; A}$ and then can perform a similar
treatment to $X^I$ and $\Psi$.
Namely, we can gauge the translation symmetry 
$b_{\mu\nu\; \ul{\alpha}} \rightarrow b_{\mu\nu\; \ul{\alpha}} +
\zeta_{\mu\nu\; \ul{\alpha}}$, where
$\ul{\alpha}=(\ul{0},\ul{a})$, by promoting $\zeta_{\mu\nu\;
  \ul{\alpha}}$ to be local and introducing the corresponding three-form gauge field $G_{\mu\nu\rho\; \ul{\alpha}}$ as
$H_{\mu\nu\rho\; \ul{\alpha}} \rightarrow 
H_{\mu\nu\rho\; \ul{\alpha}} - G_{\mu\nu\rho\; \ul{\alpha}}$.
It should be noted that the gauge transformation of $G$ is therefore
$\delta G_{\mu\nu\rho\; \ul{\alpha}} = 3 D_{[\mu} \zeta_{\nu\rho]\;
  \ul{\alpha}}$.
Then we can eliminate $H_{\mu\nu\rho\; \ul{\alpha}}$ by the gauge symmetry
and the equations of motion of $G$ gives the condition
$H_{\mu\nu\rho\; 0}=H_{\mu\nu\rho\; a}=0$.
In this way, the fields of $u^0$ and $u^a$ components
are again being moduli:
\begin{align}
  X^I_0 = \lambda_0^I \,,
\quad
  X^I_a = \lambda_a^I \,,
\quad
\Psi_0 = \Psi_a = H_{\mu\nu\rho\; 0}=H_{\mu\nu\rho\; a}=0 \,.
\end{align}
We introduce the indices $\alpha,\beta$ to represent $(0,a=1 \dots d)$
indices collectively.
Then $\lambda_\alpha^I$ are $5\times (d+1)$ matrices, and we 
define $(d+1)\times 5$  matrices $\pi_I^\alpha$ such that
\begin{align}
  \lambda_\alpha^I \pi_I^\beta = \delta^\beta_\alpha \,.
\end{align}
Such $\pi_I^\alpha$ can exist when $d \leq 4$, and we simply assume
their existence in this discussion.
Finally, we define the projector 
$P^I_J = \delta^I_J - \lambda_\alpha^I \pi^\alpha_J$ and introduce
the decomposition 
$X^I_{(i\vec{m})} = P^I_J X^J_{(i\vec{m})} 
+ \lambda^I_\alpha Y^\alpha_{(i\vec{m})}$ as before.
Here $Y^\alpha_{(i\vec{m})} = \pi^\alpha_J X^J_{(i\vec{m})}$.
Then the effective action (\ref{eq:Lag_PST_1}) becomes
\begin{align}
  \label{eq:Lag_PST_2}
  \mathcal{L}=&
-\frac{1}{2}P_{IJ} 
  \left( \tilde{D}^\mu X^I \right)_{(i,-\vec{m})}
  \left( \tilde{D}_\mu X^J \right)_{(i\vec{m})}
\nn\\ &
-\frac{1}{2}\lambda^I_\alpha \lambda^I_\beta
  \left( \tilde{D}^\mu Y^\alpha_{(i,-\vec{m})}
      - \tilde{A}^{\mu \alpha}{}_{(i,-\vec{m})} \right)
  \left( \tilde{D}_\mu Y^\beta_{(i\vec{m})}
      - \AAB{\mu} {\beta}{(i\vec{m})} \right)
\nn\\ &
+\frac{i}{2} \Bar\Psi_{(i,-\vec{m})}
 \Gamma^\mu \left( \tilde{D}_\mu \Psi \right)_{(i\vec{m})}
\nn\\ &
+\frac{1}{4} H^*_{\mu\nu\; (i,-\vec{m})}
    \left( H^{* \mu\nu} - H^{\mu\nu} \right)_{(i\vec{m})} \,.
\end{align}

\paragraph{Interpretation of the effective action}

Let us consider the brane interpretation of our effective action
(\ref{eq:Lag_PST_2}).
First note that by setting all $\lambda_\alpha^I=0$, the effective
action is, apart from the flat connection gauge field
$\AAB{\mu}{B}{A}$,
nothing but the PST Lagrangian:
\begin{align}
\label{eq:PST_Lag}
  \mathcal{L}_\text{PST} =&
- \left(\sqrt{-\det{\left( g_{\mu\nu} + i H^*_{\mu\nu} \right)}}
+\frac{1}{4} 
H^*_{\mu\nu} H^{\mu\nu}\right) \,,
\\
g_{\mu\nu} =& \partial_\mu X^M \partial_\nu X^N G_{MN} \,,
\nn\\
H_{\mu\nu} =& \frac{\partial^\rho a}{\sqrt{(\partial a)^2}}
    H_{\mu\nu\rho} \,,
\qquad
H^*_{\mu\nu} =\frac{\partial^\rho a}{\sqrt{(\partial a)^2}}
  \frac{\epsilon_{\mu\nu\rho\lambda\sigma\tau}}{3!}  H^{\lambda\sigma\tau}
\,,
\end{align}
expanded up to the quadratic order
in the flat metric with the static gauge:
\begin{align}
  G_{MN} =&
  \begin{pmatrix}
    \eta_{\mu\nu} & \\
     & \delta_{IJ}
  \end{pmatrix} \,,
\qquad
X^M = (x^\mu , X^I ) \,,
\end{align}
where $M,N= (\mu, I)$. 
For the current purpose, it is sufficient to consider only the bosonic
part of the action.
To compare it to the case with non-zero $\lambda_\alpha^I$, we
consider the following Kaluza-Klein compactification ansatz,
\begin{align}
  G_{MN} =&
  \begin{pmatrix}
    \eta_{\mu\nu}+g_{\alpha\beta} A_{\mu}{}^\alpha A_{\nu}{}^\beta 
 & - g_{\beta\gamma} A_{\mu}{}^\gamma & 0 \\
     - g_{\alpha\gamma} A_\nu{}^\gamma   & g_{\alpha\beta} & 0 \\
     0 & 0 &  \delta_{\hat{I} \hat{J}}
  \end{pmatrix} \,,
\end{align}
with the static gauge $X^M = (x^\mu , Y^\alpha , X^{\hat{I}})$.
With this ansatz, (\ref{eq:PST_Lag}) becomes, up to the quadratic
order in the physical fields:
\begin{align}
    \label{eq:PST_Lag_KK}
    \mathcal{L}_\text{PST} =&
-\frac{1}{2} \left( \partial^\mu X^{\hat{I}} \right)
             \left( \partial_\mu X^{\hat{I}} \right)
-\frac{1}{2} g_{\alpha\beta} \eta^{\mu\nu}
 \left( \partial_\mu Y^\alpha - A_\mu{}^\alpha \right)
 \left( \partial_\nu Y^\beta - A_\nu{}^\beta \right)
\nn\\ &
+\frac{1}{4} H^*_{\mu\nu}
    \left( H^{* \mu\nu} - H^{\mu\nu} \right) 
\,,
\end{align}
where we have dropped an uninteresting constant term.

We compare the resulting Lagrangian with our effective Lagrangian 
(\ref{eq:Lag_PST_2}).
Now because of the projector, some of $I,J$ directions are eliminated.
Therefore the un-eliminated index $I,J$ can be identified with
$\hat{I},\hat{J}$ here.
The projected scalars $Y^a_{(i\vec{m})}$ are identified with the
directions in which the Kaluza-Klein reduction has been performed.
The gauge fields $\AAB{\mu}{\alpha}{(i\vec{m})}$ are regarded as the
graviphoton gauge fields from the reduction.
Because of $C^{\mu}_A=0$, there is no relation between 
$\AAB{\mu}{\alpha}{(i\vec{m})}$ and $H_{\mu\nu\rho\;(i\vec{m})}$ in this
case.
It is consistent to the fact that
$ \AAB{\mu}{\alpha}{(i\vec{m})}$ is identified with an external
graviphoton.
It should be noted that since $\AAB{\mu}{\alpha}{(i\vec{m})}$ is pure gauge,
the corresponding graviphoton field should also be trivial one.
The fermions and the three-form field strength are naturally
understood.
The metric $g_{\alpha\beta}$ should be identified with
$\lambda_\alpha^I \lambda_\beta^I$
in (\ref{eq:Lag_PST_2}), and then the target space is
regarded as $\bR^{1,9-d} \times M_{d+1}$ where $M_{d+1}$ is a $d+1$
dimensional manifold with the metric $g_{\alpha\beta} =
\lambda^I_\alpha \lambda^I_\beta$. 
In the effective action, the index $(i\vec{m})$ shows that the fields
are bunch of Abelian 5-brane fields which are interacting only through
the covariant derivative.
However, as seen, the field strength of the connection vanishes, and
then these copies of the Abelian fields are very loosely communicating
each other.
Thus we have a (almost) trivially interacting Abelian fields on a five-brane
in $\bR^{1,9-d} \times M_{d+1}$.
This is a (non-Abelian) generalization of the second order NS5-brane
Lagrangian discussed in \cite{Bandos:2000az}.
In our case,  the dimensional reduction is done to  more than one
direction.


\section{Five-brane actions and duality relations}
\label{sec:5-branes}

As it has already been shown, the effective actions derived from the
equations of
motion of non-Abelian $(2,0)$ tensor multiplets in six dimensions
correspond to various brane effective actions on some torus.
Since we would like to understand the starting equations of motion
 as a kind of
effective description of multiple M5-branes, one question naturally
arises: do these effective actions respect the symmetries of
M-theory, especially string duality?

To answer this question, we analyze the effective actions in the case of 
$d=1$, namely the label for the Lorentzian generator $a$ takes only
one value $a=1$. 
This setup leads to different kinds of 5-branes,  and we will
investigate the relation between their effective actions.

\subsection{D5-branes and NS5-branes in type IIB theory}
\label{sec:5-branes2}

For the case of $d=1$, we have various five-brane actions.
We will start with 5-branes with $C_0^\mu, C_a^\mu \neq 0$ case, as in section \ref{sec:gen_case}.
The effective Lagrangian in this case is written as
\begin{align}
    \label{eq:5-brane_ac_gen}
{\cal L}_5 =&
-\frac{1}{2}
\left[
   \left( \hat{D}^\mu X^{\hat{I}} \right)_{(i,-\vec{m})}
   \left( \hat{D}_\mu X^{\hat{I}} \right)_{(i\vec{m})}
+C^2 \tau^2 
   \left( \hat{D}_{\hat{a}} X^{\hat{I}} \right)_{(i,-\vec{m})}
   \left( \hat{D}_{\hat{a}} X^{\hat{I}} \right)_{(i\vec{m})} \right]
\nn\\ &
+\frac{1}{4} C^2 
  \commt{X^{\hat{I}}}{X^{\hat{J}}}_{(i,-\vec{m})}
  \commt{X^{\hat{I}}}{X^{\hat{J}}}_{(i\vec{m})}
\nn\\ &
+\frac{i}{2} \Bar\Psi_{(i,-\vec{m})}
 \left( \Gamma^\mu ( \hat{D}_\mu \Psi )
-C \Gamma_5 \Gamma^{6} \tau ( \hat{D}_{\hat{a}} \Psi )
 \right)_{(i\vec{m})} 
\nn\\ &
+\frac{1}{2} \Bar\Psi_{(i,-\vec{m})}
 C  \Gamma_5 
 \Gamma^{\hat{I}} \commt{X^{\hat{I}}}{\Psi}_{(i\vec{m})}    
\nn\\ &
-\frac{1}{4 C^2}
\left[
F_{\mu\nu\; (i,-\vec{m})} F^{\mu\nu}_{(i\vec{m})}
+2 C^2 \tau^2 \eta_{\mu\nu}
F^{\hat{a}\mu}_{(i,-\vec{m})} F^{\hat{a}\nu}_{(i\vec{m})}
\right]
\,,
\end{align}
where $C_0^{\tilde{\mu}=5}=C$.
Since we now have only $a=1$, we can make a rotation
to set $\tau_1^I= \tau_1^6 = \tau \delta^{I6}$.
Subsequently, it is easy to see that the projector
$P^I_J = \delta^{\hat{I}}_{\hat{J}}$ where $\hat{I}, \hat{J} =
7,8,9,10$.
Namely, $P^{I}_J$ provides a projection to a plane perpendicular to the $6$ direction.
So far, we have only one internal coordinate $y$, 
corresponding to the direction which is denoted as
$\hat{a}$ hereafter.
The actual direction will be specified soon below.

This theory is compactified on a circle, whose radius is at first
considered to be constant.
The Fourier modes are expanded with the basis $e^{i y m}$,
and then $y$ becomes dimensionless.
Since the combination $C\tau$ always has mass dimension equal to $1$, we can
redefine the dimensionless coordinates $y$ and dimensionless gauge
field $A_{\hat{a}} = Y$ as
\begin{align}
    y \rightarrow C \tau y \,,
\qquad
    A_{\hat{a}} \rightarrow (C \tau)^{-1} A_{\hat{a}} \,.
\end{align}
Then the periodicity of $y$ is equal to $y \sim y + 2\pi/C\tau$, and 
the radius of the circle is identified with $R=1/(C\tau)$.

\paragraph{$D5$-brane}

Since we have started with the equations of motion, the overall factor of
the Lagrangian cannot be fixed.
Consequently, we assume a pre-factor which is equivalent to the
$D5$-brane tension $T_5 = (2\pi)^{-5} \ell_s^{-6} g_s^{-1}$,
and compare our effective action with the $D5$-brane action.
By introducing $\hat{\mu}, \hat\nu = 0,1,2,3,4, \hat{a}$,
we write the effective action as follows:
\begin{align}
\label{eq:5d_eff2}
    S_{eff}=& -T_5 \int d^5 x \sum_{m} \left(
\frac{1}{2} 
\left(\hat{D}_{\hat{\mu}} X^{\hat{I}} \right)_{(i,-\vec{m})}
\left(\hat{D}_{\hat{\mu}} X^{\hat{I}} \right)_{(i\vec{m})}
+ C^2 \tau^2 m^2 
  X^{\hat{I}}_{(i,-\vec{m})} X^{\hat{I}}_{(i\vec{m})}
\right.\nn\\&\left.\hskip1em
- \frac{1}{4} C^2 
  \commt{X^{\hat{I}}}{X^{\hat{J}}}_{(i,-\vec{m})}
  \commt{X^{\hat{I}}}{X^{\hat{J}}}_{(i\vec{m})}
+ \frac{1}{4C^2} F^{\hat{\mu}\hat\nu}_{(i,-\vec{m})}
              F_{\hat{\mu}\hat\nu\; (i\vec{m})}
+ \text{(fermions)}
\right) \,.
\end{align}
The relevant part of the action of D5-brane on $S^1$ of radius
$R$, which can be derived as
the Yang-Mills limit of the Dirac-Born-Infeld (DBI) action,
is
\begin{align}
  \label{eq:D5-action}
  S_{D5} =&
 - T_5 \int d^5 x \sum_{m} \left(
\frac{1}{2} 
\left(\hat{D}_{\hat{\mu}} X^{I} \right)_{(i,-\vec{m})}
\left(\hat{D}_{\hat{\mu}} X^{I} \right)_{(i\vec{m})}
+ \frac{m^2}{R^2} 
  X^{I}_{(i,-\vec{m})} X^{I}_{(i\vec{m})}
\right.\nn\\&\left.\hskip1em
- \frac{1}{4 (2\pi\alpha')^2}  
  \commt{X^{I}}{X^{J}}_{(i,-\vec{m})}
  \commt{X^{I}}{X^{J}}_{(i\vec{m})}
+ \frac{(2\pi\alpha')^2}{4} F^{\hat{\mu}\hat\nu}_{(i,-\vec{m})}
              F_{\hat{\mu}\hat\nu\; (i\vec{m})}
+ \cdots
\right) \,.
\end{align}
Thus we can identify
\begin{align}
\label{eq:C_alpha}
    C^2 = \frac{1}{(2\pi\alpha')^2} \,,
\qquad
R=\frac{1}{C\tau} \,,
\qquad
\tau  \propto \frac{\alpha'}{R} = R_{IIA}.
\end{align}
It should be noted that the parameter $\tau$ is related to  the radius of the T-duality circle
$R_{I{}IA}$ in the type I{}IA theory description.
Based on the identification of $\tau$,
we have KK-momentum modes along the $\tau=\tau_1^6$ direction,
and have set $\hat{a}=6$ here.
As a result, the world-volume extends on this $6$ direction as well.
This can be understood as T-duality in the specific direction.
On the other hand, because of $C_0^5$, the reduced direction $5$ is understood
as the M-theory direction,
and  we could think the $5$ direction as a circle of radius 
$R_M=g_s \ell_s$.
For this case, we can analyze the preserved supersymmetry, and the
theory has indeed non-chiral ${\cal N}=(1,1)$ supersymmetry.
See Appendix \ref{sec:(1,1)susy} for details.
Therefore, the action (\ref{eq:5d_eff2}) is the $D5$-brane action of
type I{}IB string theory.

One may wonder why this identification is so different from the
one which Lambert and Papageorgakis took in \cite{Lambert:2010wm},
where $C$ is identified as the coupling constant, and more suitable for the interpretation of the novel Higgs mechanism.
To understand the difference, we choose another coefficient to describe the effective action: 
\begin{align}
  S_5 = - \frac{1}{C^2} \int d^6 x \; \text{tr}\,
 \left[
\frac{1}{2} 
\left(\hat{D}_{\hat{\mu}} X^{I} \right)^2
-\frac{1}{4}\commt{X^I}{X^J}^2
+\frac{1}{4} F_{\mu\nu} F^{\mu\nu}
+ \dots \right] \,,
\end{align}
where $d^6 x$ includes $dy$ (of length dimension $1$)
 and $\tr$ denotes the summation over index $i$.
Since $F_{\mu\nu}$ has a  mass dimension of $2$,
$X^I$ and $C$ thus have a mass dimension of $1$ and $-1$,
respectively.
The overall factor should be identified with the Yang-Mills coupling
constant in six dimensions,
and also be proportional to the string coupling as follows:
\begin{align}
 C^2 = \left(g^{(6)}_{\text{YM}} \right)^2 = g_s \alpha' \,.
\end{align}
Upon reduction to five dimensions, this might be related to the
identification in \cite{Lambert:2010wm}.
We however preferred our previous choice because it is more
convenient to figure out another choice of the parameters for the description of NS5-branes.

\paragraph{NS5-brane}

The previous identification of the $S^1$ direction is natural, since
the covariant derivative along the internal directions has always been 
in accordance with $\tau_a^I$ as $\tau_a^I \hat{D}^a$.
Therefore, we can interpret $\tau_a^I$ as a vielbein to transform
internal directions to directions transverse to the brane.
However, for $d=1$ case, $\vec{m}$ has only one component, i.e., 
it is a number.
Furthermore, since $m$ always appears  with $C_0^5$, we can interpret $C_0^5 im
= C_0^5 \partial_y$ as a derivative along the $5$ direction, with
$C_0^5$ as a vielbein.
To see what this re-interpretation leads, we will set 
the $5$ direction as an internal direction, $\hat{a}=5$.

Once we have chosen the $5$ direction as the $S^1$ direction in which
Kaluza-Klein momentum is defined and T-duality will be taken,
we can consider the other direction, which is specified by $\tau=\tau_1^6$,
as the compactification direction for M-theory.
Subsequently, we first obtain NS5-brane in type IIA theory through the
reduction of the $6$ direction, which is transverse to the world-volume of
5-brane. Further compactification on $S^1$ in the $5$ direction
leads to T-duality, and finally we get NS5-brane in type IIB theory.
Let us look at how it works.
The radius of $S^1$ along direction $5$  is given by the expression $R=(C \tau)^{-1}$.
On the other hand,
$\tau$ has the dimension of length, and is indeed related to
the magnitude of the vanishing direction,
$\tau=\tau_1^6 = X_1^6 - v_1 X_0^6$.
Therefore, it is natural that we identify it with the radius of the
M-circle, $|\tau| =c_1 \hat{g}_s \ell_s$, where a proportional constant
$c_1$ is inserted for the sake of generality.
It should be noted that we use a different label for the string coupling, $\hat{g}_s$.
In the following subsection, this identification is more
justified in an effective theory which is related to the current model
in an Abelian limit.
These relations lead to
\begin{align}
  C =& \frac{1}{c_1 \hat{g}_s \ell_s R} 
= \frac{1}{c_1 c_2} \frac{1}{\hat{g}_s \ell_s^2} \,,
\end{align}
where $R= c_2 \ell_s$ with a constant of order  probably larger than 1 is not involved in
the string coupling.
By plugging this in (\ref{eq:5d_eff2}),
the result can be identified with (\ref{eq:D5-action}) with the
following replacements:
\begin{align}
    g_s \rightarrow \hat{g}_s = g_s^{-1} \,,
\qquad
\alpha' \rightarrow \hat{g}_s \alpha' \,.
\end{align}
These are the standard S-duality relations in type IIB string theory.
It should be noted that this change also alters the tension in front of the action
as
\begin{align}
  T_5=\frac{1}{g_s (2\pi)^5 \ell_s^6} \quad \rightarrow
\quad
\frac{1}{\hat{g}_s^2 (2\pi)^5 \ell_s^6} \,,
\end{align}
which is exactly the NS5-brane tension.
Therefore, by switching the interpretation, we have NS5-brane
effective action in type IIB string theory.
It should be noted that the S-duality transformation takes place
together with the conversion of
the interpretation of $\tau$.
In the case of D5-branes, $\tau$ is regarded
as the size of the circle of T-duality, $\tau = R_{IIA}$.
On the other hand, in the case of NS5-branes, $\tau$ is regarded as
the size of the M-circle, $\tau = R_{M}$.

Here we have identified the internal direction $\hat{a}$ with the $5$
or the $6$ directions.
We then have $D5$-branes for $\hat{a}=5$ and NS5-branes for $\hat{a}=6$
respectively.
We take either $C_0^5$ or $\tau^6$ as the
vielbein used to transform the internal direction $y$ into the physical
direction, and the direction $\hat{a}$
is determined by which of these we choose.
This dimensional reduction might be understood as a torus
compactification, and one of them is decompactified by means of KK momentum.
Changing the direction of the expansion by KK-modes corresponds to
the flip of the direction of the torus.
Therefore, we can see a similarity to the  ``9-11'' flip realization
of S-duality of type IIB string theory.

\subsection{Other five-branes and their relations}
\label{sec:brane_relation}
Next we consider the case of $C^{\mu}_0=0, C^{\mu}_a\neq 0$ 
with $d=1$, in section
\ref{sec:C0_zero}.
Since $d=1$, we have only a single non-zero $C_a^\mu$, and, by
a rotation, we can set it to $C_1^5$, while other components to vanish.
In addition, to be able to compare the resulting action with that in the previous
subsection,
we define $\AAB{\mu}{0}{(i\vec{m})}  $ as $-(\partial^5
A_{\mu})_{(i\vec{m})}$ with $\partial^5=im$,
motivated by the identification in section \ref{sec:gen_case}.
It should be noted that this is not removing tilde by taking the structure constant
off, but just a redefinition of the gauge field.
Furthermore if we  identify $Y_{(i{\vec{m}})}$ with $-A_{5(i\vec{m})}$,
the field strength $F_{5\mu \;(i\vec{m})}$ can be defined as follows:
\begin{eqnarray}
\partial_{\mu}Y_{(i\vec{m})}- \tilde{A}_{\mu\;(i\vec{m}) }^{\;\;0} 
=-\partial_{\mu}A_{5(i\vec{m})} +(\partial_5A_{\mu} )_{(i\vec{m})}
= F_{5\mu\;(i\vec{m})}.
\end{eqnarray} 
Through the redefinition of the gauge field,
we also have
$\tilde{F}_{\mu\nu}^0{}_{(i\vec{m})} =
\left(\partial_5 \left(
\partial_{\mu}A_{\nu}-\partial_{\nu}A_{\mu}
\right)\right)_{(i\vec{m})}
\equiv (\partial_5 F_{\mu\nu})_{(i\vec{m})}$.
Thus, from the first line of (\ref{eq:EOMgauge-Ca}),  we obtain the
following gauge field equation: 
\begin{eqnarray}
\partial^{\mu}F_{\mu\nu}+\lambda^2
\tilde{g}_{55}\partial^5F_{5\nu}=0.
\end{eqnarray} 
The effective Lagrangian is the free part of the Yang-Mills type Lagrangian for 5-branes:
\begin{eqnarray}
\label{eq:Lag2_Ca_1}
    {\cal L} &=&
-\frac{1}{2} P_{IJ} \left(
\left( \partial_\mu X^I \right)_{(i,-\vec{m})}
\left( \partial^\mu X^J \right)_{(i\vec{m})}
+\lambda^2 (C_1^5)^2
\left( \partial^5 X^I \right)_{(i,-\vec{m})}
\left( \partial^5 X^J \right)_{(i\vec{m})}
\right)
\nn\\ 
&&+\frac{i}{2} \Bar\Psi_{(i,-\vec{m})}
\left(
 \Gamma^\mu \partial_\mu \Psi
+  \lambda^I \Gamma^5 \Gamma^I C^5_{1} \partial_5 \Psi
\right)_{(i\vec{m})}
\nn\\ 
&&-\frac{1}{4}
\left( 
F^{\mu\nu}_{(i,-\vec{m})} F_{\mu\nu \; (i\vec{m})} 
+2 \lambda^2 (C_1^5)^2 F^{5\mu}_{(i,-\vec{m})} F_{5\mu\;
  (i\vec{m})}
\right) \,.
\end{eqnarray}
It is easy to see that this Lagrangian can be obtained by the 
free field limit $C_0^5 \rightarrow 0$ with the combination 
$v_1 C_0^5 = C_1^5$ fixed
in the generic 5-brane Lagrangian (\ref{eq:5-brane_ac_gen}).
It should be noted that we also need to rescale $F_{\mu\nu\; (i\vec{m})}$ by
$C_0^5$, and this limit also corresponds to the weak field limit.
Therefore it is natural that we end up with missing covariant
derivatives.
In this case, $\lambda$ indeed provides the size of the M-circle, as
discussed in section \ref{sec:C0_zero}. Since this case is related to the NS5-brane
case in the previous subsection with $\tau=v_1 \lambda$, the
identification developed earlier is also justified.

In section \ref{sec:vanishing_C}, we find that the case of
$C^{\mu}_A=0$ results in a second order PST type effective Lagrangian. This
is a free I{}IA NS5-brane without compactification
along the world-volume directions.
The corresponding situation here is that we only have $\lambda_0^I$
and $\lambda_1^J$ since $d=1$.
The $T^2$ compactification corresponds to taking 
only $\lambda_0^{10}$ and $\lambda_1^6$ to be nonzero.
Let us now compare this  to the $C_a\neq 0$ free I{}IB NS5-brane wrapping on
a circle in section \ref{sec:C0_zero}, the Lagrangian is
(\ref{eq:Lag2_Ca_1}).
Their relation can be understood as a T-dual relation. 
$C_a^5$ gives the size of compactification, while   the circle shrinks as
$C_a^5$ becomes gradually smaller. At $C_a^5=0$ the T-dual circle has 
infinite size, and the $C_a^5 \rightarrow 0$ I{}IB NS5-brane relates to
the $C_A=0$ I{}IA NS5-brane at this point.

Finally, we consider the following limit:
In the Lagrangian (\ref{eq:5-brane_ac_gen}),
we take $\tau \rightarrow 0$ with $C_0^5$ fixed.
Since in this limit, $v_a$ disappears from the Lagrangian and so does
$C_1^5$.
Therefore, in the limit we have the $D4$-branes effective Lagrangian
which is the same as the one studied in \cite{Lambert:2010wm}.
Note that since the radius of the circle is given by 
$R=1/\sqrt{C^2\tau^2}$, this is the decompactification limit in the I{}IB
side, and then we have double-dimensionally reduced $D4$-branes in I{}IA
here.
We also remark that if the limit $v_1\rightarrow 0$
in (\ref{eq:5-brane_ac_gen}), the Lagrangian remains essentially the
same, but replacing $\tau^6$ with $\lambda_1^6$.
This describes $D5$-branes in I{}IB, and by comparing  to
NS5-branes in IIB (\ref{eq:Lag2_Ca_1}) we see that the role of the
moduli is just switched, namely $(C_0^5, \lambda_1^6)$ for $D5$
and $(C_1^5 , \lambda_0^6)$ for NS5.
This also resembles the ``9-11'' flip realization of S-duality in type IIB
string theory.


\section{Conclusion and Discussion}
\label{sec:conclusion}

In this paper, we examined the equations of motion proposed by Lambert
and Papageorgakis for non-Abelian $(2,0)$ tensor multiplets in six
dimensions \cite{Lambert:2010wm}.
Some of these equations are regarded as constraint equations 
for non-dynamical fields, $C^\mu_A$. 
We consider various cases where different
components of $C^\mu_A$ take non-zero values.
With respect to the cases, which components of $C^\mu_A$ 
are zero, we derive various 
equations of motion for $p$-branes 
on a backgrounds consisting of a flat space and
a torus.
We found that, in order to maintain a non-Abelian interaction after
taking into account possible 
constraints, the $u^0$ component of the $C$ field has to
be non-zero.
Otherwise, we have equations of motion of Abelian fields which are loosely
bound.
Of particular interest is the case where $d=1$,
with $d$ being the dimension of the torus. In this case,
we have a circle and 5-branes. 
For the generic $C$ case, we have Yang-Mills type actions, and
by identifying the appropriate parameters, 
we have the description of either a $D5$-brane or
NS5-brane.
In the case of 
type I{}IB string theory, the S-duality of 5-branes 
can be interpreted as the interchange of roles between two moduli fields,
which specify the compactified circles of M-theory.
We therefore observe that the formulation of non-Abelian tensor
multiplets seems to be compatible with the expectation
from the string duality.
In contrast, if only $C_a \neq 0$,
we obtain the Lagrangian for Abelian 5-branes; 
this corresponds to the free field limit of the
previous case.
We also found that the case of zero $C$ corresponds to the second order
PST-type 5-brane action.
It is worth noting that the almost free Abelian theory still includes
covariant derivatives with flat connections.
The second order PST-type action can also be considered to correspond to 
a limit of the previous two cases 
and be compatible with the expected T-duality.

In the following paragraphs,
we will discuss the findings of this work in detail 
and propose possible directions for future research.

\paragraph{Non-Abelian multiple 5-branes with (2,0) supersymmetry?}

Through the present study, we found multiple 5-branes with $(1,1)$-type
and $(2,0)$-type world-volume supersymmetries.
More specifically, 
5-branes with $(1,1)$ supersymmetry were described for 
the case of $C_0\neq 0$ by means of ordinary non-Abelian SYM.
In the case of the type IIB string theory,
they were described as D/NS 5-branes. 
On the other hand, in the case of the M-theory and the 
type IIA string theory, 5-branes are characterized by $(2,0)$-type 
world-volume supersymmetry. 
These are identified to be the 5-branes corresponding to the $C_A=0$ case.
These $(2,0)$-type 5-branes are also characterized by 
non-Abelian interactions under gauge fields. 
However, as we have shown in previous sections, 
this does not mean that our results are satisfactory.  
Despite the ``non-Abelian extension'' using the 3-algebra, these 
$(2,0)$-type multiple 5-branes have trivial non-Abelian interactions.

Indeed, there is a no-go theorem proposed in papers
\cite{Bekaert:1999dp}, claiming 
that it is impossible to obtain the 
desired non-Abelian extension of the
Abelian chiral 2-form theory in six dimensions using local
deformation terms
\footnote{This theorem was claimed to the theory of chiral 2-form
  without other fields.
On the other hand, the authors of \cite{Huang:2010rn} found that there is
no S-matrix of the (weakly-)interacting (2,0) tensor multiplet
provided that
we assume it to have a Lagrangian description.}.
In the light of this, we think one possible way of overcoming 
the problem through the introduction of 
non-local deformation of the supersymmetry
algebra (\ref{SUSY_X})
\footnote{
Some works explored  to introduce non-locality by considering loop
space variables \cite{Huang:2010db}.},
or alternatively, we may consider a 
quantization of the Nambu bracket~\cite{Dito:1996xr} 
\footnote{
In \cite{Chu:2010eb},
the authors propose a criterion which any valid quantization
of the generalized Takhtajian theory should satisfy.
It is very interesting to check that the quantization in 
\cite{Dito:1996xr} satisfies the criterion or not.}
to introduce an interesting interaction. 
This problem still remains open.

We also pose the question about the $N^3$ entropy scaling law of an $N$
M5-brane system \cite{Klebanov:1996un}.
We recall that the (truncated-)Nambu-Poisson algebra has been used to 
provide a 3-algebraic explanation for the $N^{3/2}$ entropy scaling law of an
$N$ coincident M2-brane system \cite{Chu:2008qv}.
This algebra may provide the means of explaining the
M5-branes entropy.\\

\paragraph{5D MSYM}

Finally, we give a remark that our findings are related 
to recent research on the 5D MSYM theory \cite{Lambert:2010iw}\cite{Douglas:2010iu}.
The authors of these papers discuss 
the relation between 5D MSYM and 6D $(2,0)$ superconformal field
theory on $S^1$.
More specifically, 
KK-modes of the $(2,0)$ theory, associated with $S^1$
compactification, can be explained as solitonic states in 5D MSYM.

In contrast, in our study, the $(2,0)$ theory was employed to
describe $S^1$ compactifications of the 6D theory in a number of 
special cases. 
The KK-momentum was provided by 3-algebra and independent
of the degrees of freedom of 5D SYM.
Therefore, to explain KK-modes 
in our analysis
as the solitonic states in 5D MSYM,
some non-trivial relations in addition to the
field equations (\ref{feq1})-(\ref{feq6}) are necessary.

\subsubsection*{Acknowledgments}

We would like to thank Chien-Ho Chen, Kazuyuki Furuuchi, Sheng-Lan Ko,
Yoshinori Honma, Pei-Ming Ho, Yu-Ting Huang, Eibun Senaha, Shotaro Shiba and 
Seiji Terashima
for various discussion and comments.
We also thank Yoshinori Honma, Morirou Ogawa and Shotaro
Shiba for informing us that their work \cite{HOS} significantly overlaps
ours.
TT would like to 
thank Center for Quantum Spacetime (CQUeST), 
in particular Jeoung-Hyuck Park, 
for the hospitality and support during his stay in February 2011.
The work of SK is partly
supported by Taiwan's NSC grant 097-2811-M-003-012 and
97-2112-M-003-003-MY3.
And the work of TT is 
supported by Taiwan's NSC grant 99-2811-M-002-197.
We also thank the support of NCTS.

\appendix
\section{Summary for notation}

\subsection{A Lorentzian 3-algebra}
\label{sec:algebra}

We use the loop extension of the Lorentzian 3-algebra
which is also used in \cite{Kobo:2009gz}, but with a slightly
different notation.
The generators are collectively denoted as
\begin{align}
T^A =&
\left\{ T^{(i \vec{m})} ,
u^0, u^a, u^{\ul{0}}, u^{\ul{a}} \right\} \,,
\end{align}
namely, $A$ denotes all the indices collectively, $A=\{ (i\vec{m}),
0,a,\ul{0},\ul{a} \}$.
We sometimes use the index $\alpha=(0,a)$ and
$\ul{\alpha}=(\ul{0},\ul{a})$.
We assume that our 3-algebra is equipped with
the gauge invariant metric $g_{AB}$
\begin{align}
    g_{(i\vec{m})(j\vec{n})} =& \delta_{ij} \delta_{\vec{m}+\vec{n}}
    \,,
\qquad
g_{0 \ul{0}} = 1 \,,
\qquad
g_{a \ul{b}} = \delta_{ab} \,,
\end{align}
and the other components are zero.
Here we conventionally define the generator $u_{a}$ representing the 
center element $u^{\underline{a}}$ obtained by the metric as
\begin{equation}
u_{a} \equiv g_{a\underline{b}} u^{\underline{b}} = u^{\underline{a}}.
\end{equation}
Therefore the inner products are 
\begin{align}
    \langle T^{(i\vec{m})} , T^{(i\vec{n})} \rangle
=& \delta^{ij} \delta^{\vec{m}+\vec{n}} \,,
\qquad
\langle u^a , u_b \rangle =
\langle u^a , u^{\underline{c}}g_{\underline{c}b} \rangle =
g^{a\underline{c}}g_{\underline{c}b}
= \delta^a_{\ph{a}b} \,, \nn \\
\langle u^0 , u_0 \rangle =&
\langle u^0 , u^{\underline{0}} \rangle 
= 1 \,,
\end{align}
and otherwise zero.

The structure constant of the 3-algebra is essentially the same as
the one in \cite{Kobo:2009gz}.
The nonzero components of the totally antisymmetric structure constant
are
\begin{align}
    f^{0a(i\vec{m})(j\vec{n})} =& -i m^a \delta^{ij}
    \delta^{\vec{m}+\vec{n}} \,,
\qquad
f^{0 (i\vec{m})(j\vec{n})(k\vec{\ell})} =
f^{ijk} \delta^{\vec{m}+\vec{n}+\vec{\ell}} \,,
\end{align}
which satisfy the usual fundamental identity,
\begin{align}
  \label{eq:fund_id}
  f^{ABC}{}_F f^{FDE}{}_G
+ f^{ABD}{}_F f^{CFE}{}_G
+ f^{ABE}{}_F f^{CDF}{}_G
=& f^{CDE}{}_F f^{ABF}{}_G \,.
\end{align}
The generators $u^{\ul{0}}$ and $u^{\ul{a}}$ are center, in the sense
that they do not appear as the upper index of the structure constant,
namely, when they are put inside the three-bracket defined below,
the result is zero.
On the other hand, the generators $u^0$ and $u^a$ are not in the lower
indices of the structure constant, and then they do not show up as a
result of the three-bracket operation.

We sometimes use the three-bracket expression,
\begin{align}
    \thcommt{u^0}{u^a}{u^b} =& 0 \,,
\\
    \thcommt{u^0}{u^a}{T^{(i\vec{m})}} =& m^a T^{(i\vec{m})} \,,
\\
    \thcommt{u^0}{T^{(i\vec{m})}}{T^{(j\vec{n})}} =& 
m^a \delta^{ij} \delta^{\vec{m}+\vec{n}} u_a
+i \fijk{i}{j}{k} T^{(k,\vec{m}+\vec{n})} \,,
\\
    \thcommt{u^a}{T^{(i\vec{m})}}{T^{(j\vec{n})}} =& 
- m^a \delta^{ij} \delta^{\vec{m}+\vec{n}} u^{\ul{0}}\,,
\\
    \thcommt{T^{(i\vec{m})}}{T^{(j\vec{n})}}{T^{(k\vec{\ell})}} =& 
- i f^{ijk} \delta^{\vec{m}+\vec{n}+\vec{\ell}} u^{\ul{0}} \,,
\end{align}
where $\thcommt{T^A}{T^B}{T^C} = i \fABCD{A}{B}{C}{D} T^D$.

The indices are normally used in the following conventions:
\begin{itemize}
      \item $I,J,\cdots$: the transverse directions of five branes
        ($I,J=6,\cdots,10$).
      \item $\mu,\nu,\cdots$: world-volume directions of five branes
    ($\mu,\nu=0,\cdots,5$).
      \item $d$: The number of the Lorentzian generators($-1$):
   $a , \underline{a}=1,\cdots,d$.
     \item $y_a$: The coordinates for the torus $T^d$.
The Fourier basis along the torus is $e^{i m^a y_a}$
and $\partial^a = \frac{\partial}{\partial y_a}$.
We will regard the $T^{(i\vec{m})}$ components of the fields,
$\phi_{(i\vec{m})}$, as the Fourier components of the field in $y$
coordinates,
\begin{align}
    \phi(x,y) =& \sum_{\vec{m}} \phi_{(i\vec{m})}(x) e^{i \vec{m}
      \cdot \vec{y}} \,,
\end{align}
and also use the expression $i m^a \phi_{(i\vec{m})} = 
\left( \partial^a \phi \right)_{(i\vec{m})}$.
\end{itemize}

\subsection{Covariant derivative, field strength}
\label{sec:DandF}

Our convention for the covariant derivative and the field strength is
the same as in \cite{Lambert:2010wm}.
\begin{itemize}
\item Covariant derivative:
$(D_{\mu}\phi)_A=\partial_{\mu}\phi_A-\AAB{\mu}{B}{A}\phi_B$

\item Field strength:
$ \FAB{\mu}{\nu}{B}{A}= -\partial_{\mu}\AAB{\nu}{B}{A}+\partial_{\nu}\AAB{\mu}{B}{A}
+\AAB{\mu}{C}{A}\AAB{\nu}{B}{C}-\AAB{\nu}{C}{A}\AAB{\mu}{B}{C}$

\item Covariant derivative of field strength:

 $ \left(D^\mu \tilde{F}_{\mu\nu} \right)^B{}_A
=
\partial^\mu \FAB{\mu}{\nu}{B}{A}
+\tilde{A}^{\mu\, B}{}_C \FAB{\mu}{\nu}{C}{A}
-\tilde{A}^{\mu\, C}{}_A \FAB{\mu}{\nu}{B}{C}
$
\end{itemize}

\subsection{Assumption for the gauge field}
\label{sec:assum_gauge}

We assume that the
gauge fields $\tilde{A}_\mu$ are accompanied with the structure
constant of 3-algebra,
\begin{align}
    \tilde{A}^B_{\mu \, A} =&A_{\mu\, CD} f^{CDB}_{\ph{CDB}A} \,,
\end{align}
which guarantees that it acts on the three-bracket as a derivation.
Due to the limited form of the structure constant, we have
\begin{align}
\label{eq:vanishing_A}
\AAB{\mu}{0}{\ul{0}}=&
\AAB{\mu}{a}{\ul{b}}=
\AAB{\mu}{A}{0}=
\AAB{\mu}{A}{a}=
\AAB{\mu}{\ul{0}}{A}=
\AAB{\mu}{\ul{a}}{A}=0 \,.
\end{align}
The rest can be written in terms of the gauge field without tilde,
\begin{align}
\label{tilA_ntilA1}
    \AAB{\mu}{0}{(i\vec{m})} =&
\fijk{j}{k}{i} A_{\mu \, (j\vec{n})(k,\vec{m}-\vec{n})}
-i m^a A_{\mu \, a (i\vec{m})}
\nn\\=&
\fijk{j}{k}{i} A_{\mu \, (j\vec{n})(k,\vec{m}-\vec{n})}
-(\partial^a A_{\mu \, a})_{(i\vec{m})} \,,
\\
\label{tilA_ntilA2}
    \AAB{\mu}{a}{(i\vec{m})} =&
i m^a A_{\mu \, 0 (i\vec{m})}
=
(\partial^a A_{\mu \, 0})_{(i\vec{m})} \,,
\\
\label{tilA_ntilA3}
    \AAB{\mu}{0}{\ul{a}} =&
- i m^a A_{\mu \, (i\vec{m})(i,-\vec{m})} \,,
\\
\label{tilA_ntilA4}
    \AAB{\mu}{(i\vec{m})}{(i\vec{m})} =&
- i m^a A_{\mu \, 0 a} \,,
\\
\label{tilA_ntilA5}
    \AAB{\mu}{(i\vec{m})}{(j\vec{n})} =&
\fijk{k}{i}{j} A_{\mu \, 0 (k,\vec{n}-\vec{m})} \,.
\end{align}

Since some components of the gauge field are zero by 3-algebra
as summarized above,
the covariant derivatives take the following form before further
restriction,
\begin{align}
    \left( D_\mu \phi \right)_{(i\vec{m})} =& 
\left( \tilde{D}_\mu \phi \right)_{(i\vec{m})} 
- \tilde{A}^0_{\mu\, (i\vec{m})} \phi_0
- \tilde{A}^a_{\mu\, (i\vec{m})} \phi_a \,,
\\
\label{eq:def_Tilde_D}
\left( \tilde{D}_\mu \phi \right)_{(i\vec{m})}  =&
\partial_\mu \phi_{(i\vec{m})}
-\tilde{A}^{(i\vec{m})}_{\mu\, (i\vec{m})} \phi_{(i\vec{m})} 
-\tilde{A}^{(j\vec{n})}_{\mu\, (i\vec{m})} \phi_{(j\vec{n})} \,,
\\
\label{eq:def_Hat_D}
\left( \hat{D}_\mu \phi \right)_{(i\vec{m})}  =&
\partial_\mu \phi_{(i\vec{m})}
-\tilde{A}^{(j\vec{n})}_{\mu\, (i\vec{m})} \phi_{(j\vec{n})} 
= \partial_\mu \phi_{(i\vec{m})}
+i[{A}_{\mu}, \phi]_{(i\vec{m})} 
\end{align}
where $\phi_A$ denotes collectively the physical fields $X^I_A$,
$\Psi_A$ and $H_{\mu\nu\rho\; A}$.

\subsection{Anti-symmetrization}

We use the following convention for the totally anti-symmetric
combination of the indices:
\begin{align}
    A^{[\mu} B^{\nu]} =& \frac{1}{2} \left(A^\mu B^\nu - A^\nu B^\mu \right)
\,,
\\
    A^{[\mu} B^{\nu} C^{\rho]} =& \frac{1}{3!} \left(A^\mu B^\nu
      C^\rho - A^\nu B^\mu C^\rho
+\text{(4 other terms)}
\right) \,,
\end{align}
namely, in general,
\begin{align}
    A_1^{[\mu_1} A_2^{\mu_2} \cdots A_n^{\mu_n]}
=&
\frac{1}{n!} \left(
A_1^{\mu_1} A_2^{\mu_2} \cdots A_n^{\mu_n}
-A_1^{\mu_2} A_2^{\mu_1} \cdots A_n^{\mu_n}
+\cdots
\right)  \,,
\end{align}
where $\cdots$ denotes the all possible combination of the indices
 with sign.

So, for example,
\begin{align}
    D_{[\mu} H_{\nu\rho\sigma]} =& 
\frac{1}{4} \left(
D_\mu H_{\nu\rho\sigma}
-D_\nu H_{\rho\sigma\mu}
+D_\rho H_{\sigma\mu\nu}
-D_\sigma H_{\mu\nu\rho}
\right) \,.
\end{align}


\section{ ${\cal N}=(1,1)$ supersymmetry }
\label{sec:(1,1)susy}
We discuss how ${\cal N}=(1,1)$ supersymmetry of D5-brane  
in section \ref{sec:5-branes} is realized
from ${\cal N}=(2,0)$ supersymmetric set up.\\

We use the same convention as in \cite{Lambert:2010wm} for Gamma matrices. 
Supersymmetry is parametrized by 16-component spinor $\epsilon$. 
The chirality of $\epsilon$ is described by 
\begin{eqnarray}
\Gamma^{012345} \epsilon
=\epsilon.
\label{susycond1}
\end{eqnarray} 

In section 
\ref{sec:5-branes}, the 5th direction of the D5-brane world volume is
reduced by the consequence of (\ref{feq6}), and other world volume
direction, say $a=6$, is created by KK momentum. Thus  the natural
chirality operator for this 5-brane is $\Gamma^{012346}$, instead of
$\Gamma^{012345}$.
Because of $\{ \Gamma^{012345}, \Gamma^{012346} \} = 0$,
$\epsilon$ in (\ref{susycond1}) contains both chirality states of
$\Gamma^{012346}$.
If we take a basis of the spinor 
\begin{eqnarray}
\epsilon= \left(\begin{array}{c}
\epsilon_{+} \\ \epsilon_{-}
\end{array}\right), \qquad 
\Gamma^{012346}\epsilon_{\pm}=\pm\epsilon_{\pm},
\end{eqnarray}
 $\Gamma^{012345}$ can be written by a $16\times 16$ matrix $\gamma^{012345}$  
\begin{eqnarray}
\Gamma^{012345}=
\left(\begin{array}{cc}
 0 & \gamma^{012345} \\ 
\gamma^{012345} & 0 
\end{array} \right). 
\end{eqnarray} 
The supersymmetric condition (\ref{susycond1}) gives
\begin{eqnarray}
 \gamma^{012345}\epsilon_{\pm}=\epsilon_{\mp}.
\end{eqnarray}
Since $\gamma^{012345}$ is invertible,
the numbers of the components of $\epsilon_{\pm}$ are the same.
They describe the vector-like $(1,1)$ supersymmetry in six dimensions,
as expected for D5-brane.


\section{The relationship between $a^{\mu}$ and the 2-form gauge field 
$b_{\mu\nu}$}
\label{sec:amu and 2-form}

From (\ref{feq2}), it is easy to see that
for $H_{\mu\nu\rho\; \alpha}$ 
the equations of motion are the usual Bianchi identity
$\partial_{[\mu}H_{\nu\rho\sigma]\,\alpha} = 0$,
and then these
3-form field strength can be represented by the 2-form fields
$b_{\mu\nu\; \alpha}$,
\begin{align}
  H_{\nu\rho\sigma\,a} = 
\partial_{\nu} b_{\rho\sigma\,\alpha}
+\partial_{\rho} b_{\sigma\nu\,\alpha}
+\partial_{\sigma} b_{\nu\rho\,\alpha} \,.
\end{align}
From (\ref{feq4}) and the relation (\ref{tilA_ntilA4}),
one can see that
the field strength $f_{\mu\nu\; a}$ of the gauge field
$a_{\mu\; a} \equiv A_{\mu\; a0}$
is represented by the 3-form field strength as
\begin{equation}
m^b f_{\mu\nu\; b} = 
-
m^a\left(
C^{\rho}_{0} H_{\mu\nu\rho\; a}
-C_a^\rho H_{\mu\nu\rho\; 0} \right), 
\label{delf0}
\end{equation}
where $f_{\mu\nu \;a } 
= ( \partial_{\mu}a_{\nu\; a}
- \partial_{\nu}a_{\mu\; a})$.
The constraint (\ref{feq6}) causes the dimensional reduction of the
three-form field,
and therefore we take the two-form potential to obey the
relation,
\begin{equation}
C^{\rho}_{0}\partial_{\rho} b_{\mu\nu\,a}
- C^{\rho}_{a}\partial_{\rho} b_{\mu\nu\,0} = 0 \,,
\end{equation}
as well as the one-form gauge transformation as we will see soon.
We can then identify
\begin{equation}
a_{\mu\,a} = 
- C^{\nu}_{0} b_{\mu\nu\; a} 
+ C_a^\nu b_{\mu\nu\; 0} \,.\label{2-form-1-form}
\end{equation}

Under the dimensional reduction, 
the $U(1)$ gauge transformation of the 
2-form gauge fields $b_{\mu\nu\,\alpha}$,
\begin{align}
b_{\mu\nu \, \alpha} \to b_{\mu\nu\,\alpha}
+ \delta b_{\mu\nu\,\alpha}
= b_{\mu\nu\,\alpha}
+ \partial_{\mu}\Lambda'_{\nu\,\alpha}
-\partial_{\nu}\Lambda'_{\mu\,\alpha} \,,
\end{align}
is naturally identified with 
the $U(1)$ gauge transformation of $a_{\mu\,a}
\to a_{\mu\,a} + \partial_{\mu}\Lambda_{a}$
through the identification (\ref{2-form-1-form}) 
\footnote{The $U(1)$ gauge transformation is generated by
the gauge transformation parameter with the Lie 3-algebra
\begin{equation}
\tilde{\Lambda}^{(i\vec{m})}{}_{(i\vec{m})} = 
-i m_a\Lambda_{0a} =
i m_a\Lambda_{a} \,.
\end{equation}}.
Through the (\ref{2-form-1-form}), 
the $U(1)$ gauge transformation of the 2-form gauge field
generate the gauge transformation of the 1-form gauge field as
\begin{align}
\delta a_{\mu\,a} =&
- C^{\nu}_{0} \delta b_{\mu\nu\,a} 
+ C^{\nu}_{a} \delta b_{\mu\nu\,0}\, \nn \\ 
=&
- 2C^{\nu}_{0} \partial_{[\mu} \Lambda'_{\nu]\,a} 
+ 2C^{\nu}_{a} \partial_{[\mu} \Lambda'_{\nu]\,0} \, \nn \\
=&
- C^{\nu}_{0} \partial_{\mu} \Lambda'_{\nu\,a} 
+ C^{\nu}_{a} \partial_{\mu} \Lambda'_{\nu\,0} 
\end{align}
due to the dimensional reduction 
$C^{\nu}_{0} \partial_{\nu} \Lambda'_{\mu\; a} 
- C^{\nu}_{a} \partial_{\nu} \Lambda'_{\mu\; 0} = 0$.
Finally the $U(1)$ gauge transformation of the 1-form gauge field
can be unified to the $U(1)$ gauge transformation of the 2-form gauge field
through the identification of the parameter
\begin{equation}
\Lambda_{a} =
- C^{\nu}_{0} \Lambda'_{\nu\,a} 
+ C_a^\nu \Lambda'_{\nu\,0}
\,.
\end{equation}

This relation is interesting since, at the beginning, the 3-algebra
gauge transformation and the two-form gauge field transformation are
the different things, but now they are unified\footnote{%
Only a  part of 2-form gauge transformation is related to 1-form one. 
2-form gauge transformations that are not related to  $C^{\mu}$
directions are invisible by the 1-form $a_{\mu}$.}.
If it also worked in the non-Abelian part, this mechanism would give a
significant
suggestion for the non-Abelian generalization of two-form gauge field
(or maybe higher form even).
Unfortunately, it seems that this
unification can be confirmed only in this Abelian sector.

Since the three-form field strength $H_{\mu\nu\rho\; \alpha}$ is
self-dual, the Bianchi identity implies that it satisfies the usual
equation of motion without sources,
\begin{align}
  \partial^\mu H_{\mu\nu\rho\; \alpha} =0 \,.
\end{align}
This suggests that $m^a \partial^\mu f_{\mu\nu \; a}=0$
for arbitrary $\vec{m}$, and then $\partial^\mu f_{\mu\nu\; a}=0$.
However the rest of the dynamical fields, $X^I_{(i\vec{m})}$,
$\Psi_{(i\vec{m})}$ and so on, couple to $a_{\mu\; a}$ through the
covariant derivative, and then they will generate the source term for
this field strength.
In order for this equation of motion to hold, we need to regard
$a_{\mu\; a}$ as the background field and we do not take the variation
with respect to $a_{\mu \; a}$ in the other part of the Lagrangian.


\end{document}